\documentstyle{mn}
\def\eg{{\it e.g. }}
\def\etal{{\it et al. }}
\def\deg{\ifmmode^\circ\else$^\circ$\fi}
\def\ltsima{$\; \buildrel < \over \sim \;$}
\def\simlt{\lower.5ex\hbox{\ltsima}}
\def\gtsima{$\; \buildrel > \over \sim \;$}
\def\simgt{\lower.5ex\hbox{\gtsima}}
\def\hmpc{h^{-1}{\rm Mpc}}

\def\muK{\mu{\rm K}}

\newcounter{parentequation}
\def\beglet{
  \addtocounter{equation}{1}%
  \setcounter{parentequation}{\value{equation}}%
  \setcounter{equation}{0}%
  \def\theequation{\arabic{parentequation}\alph{equation}}%
  \ignorespaces
}
\def\endlet{
  \setcounter{equation}{\value{parentequation}}%
  \def\theequation{\arabic{equation}}%
}
\begin{document}

\title[Cosmological Parameters from CMB anisotropies]{Cosmic Confusion: 
Degeneracies among\\
Cosmological Parameters Derived from \\
Measurements of Microwave Background Anisotropies}

\author[G. Efstathiou and J.R. Bond]{G. Efstathiou$^{1}$ and  J.R. Bond$^{2}$\\
$^1$ Institute of Astronomy, Madingley Road, Cambridge, CB3 OHA.\\
$^2$ CITA, 60 St. George St., University of Toronto, Toronto ON M5S
3H8, Canada.\\}

\maketitle

\begin{abstract}
In the near future, observations of the cosmic microwave background
(CMB) anisotropies will provide accurate determinations of many
fundamental cosmological parameters. In this paper, we analyse
degeneracies among cosmological parameters to illustrate some of the
limitations inherent in CMB parameter estimation. For simplicity,
throughout our analysis we assume a cold dark matter universe with
power-law adiabatic scalar and tensor fluctuation spectra.  We show
that most of the variance in cosmological parameter estimates is
contributed by a small number (two or three) principal components. An
exact likelihood analysis shows that the usual Fisher matrix approach
can significantly overestimate the errors on cosmological parameters.
We show that small correlated errors in estimates of the CMB power
spectrum at levels well below the cosmic variance limits, (caused, for
example, by Galactic foregrounds or scanning errors) can lead to
significant biases in cosmological parameters.  Estimates of
cosmological parameters can be improved very significantly by applying
theoretical restrictions to the tensor component
and external constraints derived from 
more conventional astronomical observations such as measurements
of he Hubble constant, Type 1a supernovae distances and observations
of galaxy clustering and peculiar velocities.
\end{abstract}

%\keywords{Cosmology; Cosmic microwave background}

\section{Introduction}\label{sec:intro}

A number of investigations have shown that high precision measurements of
the cosmic microwave background (CMB) anisotropies can be used to
determine  many cosmological parameters to unprecedented precision
(Jungman \etal 1996; Bond, Efstathiou and Tegmark 1997, hereafter BET97;
Zaldarriaga, Spergel and Seljak 1997, hereafter ZSS97).  These parameters
include the amplitudes and spectral indices of the scalar and tensor
fluctuations predicted by inflationary models (\eg Knox and Turner 1994,
Knox 1995, Lidsey \etal 1997, Souradeep \etal 1998, Copeland \etal 1998), the densities of the
various components of the mass density of the Universe, and constraints on
the geometry of the Universe.

 The prospect of measuring such cosmological parameters to high
precision has formed an important part of the scientific case for two
approved satellite missions, NASA's MAP satellite (Bennett \etal 1996)
and ESA's Planck Surveyor\footnote{Formerly known as
COBRAS/SAMBA}(Bersanelli \etal 1996).  In addition, ground based and
balloon experiments are already beginning to set useful constraints on
some of these parameters ({\it e.g.} Bond and Jaffe 1997, Hancock
\etal 1997, Lineweaver and Barbosa 1998a,b, Webster \etal 1998).

In this paper, we investigate degeneracies among cosmological
parameters, {\it i.e.} to what extent can we disentangle one parameter from 
another using measurements of the CMB anisotropies alone (see
also Bond \etal 1994). We assume throughout that the primordial
fluctuations are adiabatic and Gaussian, as expected in most
inflationary models ({\it e.g.} Linde 1990). These assumptions are
physically well motivated and allow accurate calculations of the CMB
anisotropies as a function of cosmological parameters via fast
numerical solutions of the relativistic Boltzmann equation.  The
accuracy of computations of CMB anisotropies in topological defect
models has greatly improved recently (see Turok, Pen and Seljak, 1997,
Albrecht, Battye and Robinson 1997) and tend to disfavour the simplest
versions of defect models. However, these calculations cannot yet be
done to sufficient precision for the type of analysis presented in
this paper.  Nevertheless, some qualitative features of the discussion
presented here are likely to be applicable to defect models and to
other classes of models such as isocurvature theories.

We distinguish between a nearly exact degeneracy (which we call the
`geometrical' degeneracy) and other parameter degeneracies.  The
geometrical degeneracy (BET97, ZSS97) leads to near identical CMB
anisotropies in universes with different background geometries but
identical matter content. In linear perturbation theory, observations
of the primary CMB anisotropies cannot break the geometrical
degeneracy, no matter how precise the experimental measurements. The
geometrical degeneracy thus imposes fundamental limits on measurements
of the curvature of the Universe and the Hubble constant derived from
CMB anisotropy measurements.  Of the remaining parameter degeneracies,
some are extremely sensitive to the accuracy of the CMB anisotropy
measurements whilst others are not. In this paper, we give a physical
description of the causes of some of the parameter degeneracies and of
the fundamental limits imposed by CMB anisotropy measurements.

This paper is laid out as follows. In Section \ref{sec:2} we summarise
the formalism based on the Fisher matrix (Jungman \etal 1996, Tegmark,
Taylor and Heavens 1997) which is assumed in most discussions of
parameter estimation from the CMB. Throughout this paper, we use
approximate experimental parameters for the MAP and Planck satellites
to provide numerical examples of what might be achieved in the next
decade or so.  The experimental parameters that we use are summarised
in Section \ref{sec:2}. In Section \ref{sec:3}, we discuss the
geometrical degeneracy and we show that with an appropriate change of
variables, the Fisher matrix can be applied to give accurate results
for cosmological models of arbitrary curvature. In Section
\ref{sec:4}, we specialise to spatially flat cosmological models and
analyse degeneracies among other cosmological parameters. We show that
a principal component analysis provides a powerful technique for
analysing degeneracies among parameters. For experiments such as MAP
and Planck, that sample the CMB anisotropy power spectrum beyond the
first Doppler peak, the first few principal components dominate the
expected variances of most cosmological parameters.  We test the
accuracy of the Fisher matrix approach by performing an exact
likelihood analysis in a subspace defined by the two most poorly
determined principal components. Some of the degeneracies among
cosmological parameters can be understood in terms of the locations of
the Doppler peaks and of the height of the first Doppler peak. In
Section \ref{sec:4}, we compare the accuracies of parameter estimates from 
the improved version of MAP with those from Planck.  We
distinguish between `strength' and `statistical significance' in the
analysis of CMB power spectra and discuss the effects of possible
systematic errors that might arise, for example, through inaccurate
subtraction of point sources, Galactic foregrounds and map
reconstruction errors.

Section \ref{sec:5} discusses briefly how some of the degeneracies
among cosmological parameters can be removed by using other types of
observation, {\it e.g.} measurements of the Hubble constant, the age
of the Universe and constraints on the luminosity distance from
observations of distant Type 1a supernovae.  Our conclusions are
summarised in Section \ref{sec:6}.  Throughout this paper we use CMB
power spectra computed with the CMBFAST code developed by Seljak and
Zaldarriaga (1996).

\section{Fisher Matrix and Experimental Parameters}\label{sec:2}

We assume that the initial fluctuations are adiabatic and Gaussian. The
temperature anisotropies $\Delta T/T$ of the CMB on the celestial sphere
can be decomposed in a spherical harmonic expansion,
\begin{equation}
{\Delta T \over T} = \sum_{\ell, m} a_{\ell, m} Y_\ell^m(
\theta, \phi)\, ,  \label{eq:1}
\end{equation}
where each coefficient $a_{\ell, m}$ is statistically independent and
Gaussian distributed with zero mean and variance
\begin{equation}
C_\ell = \langle \vert a_{\ell, m} \vert ^2 \rangle\, .\label{eq:2}
\end{equation}
The power spectrum,  $C_\ell$, completely specifies the statistical
properties of the primary CMB temperature fluctuations if the initial
fluctuations are Gaussian. The anisotropy of Thomson scattering results in
a small net linear polarization of the CMB anisotropies (Kaiser 1983, Bond
and Efstathiou 1984).  For near scale-invariant spectra, this linear
polarization amounts to a few percent of the total temperature anisotropy
and  should be measurable in future CMB experiments such as MAP and
Planck.  Polarization introduces three  additional power-spectra to fully
characterise the temperature anisotropies.  These define the `electric'
and `magnetic' components of the polarization pattern and the cross
correlation of the polarization pattern with the temperature anisotropies
(see e.g.  Seljak and Zaldarriaga 1996, Hu and White 1997).  We do not
analyse polarization in any detail in this paper.  The effects of polarization
measurements on cosmological parameters has been discussed recently by
ZSS97.  Their results show that polarization measurements can improve the
accuracy of most cosmological parameters by relatively modest factors,
with two important exceptions, distinguishing between scalar and tensor
modes and determining the redshift at which the intergalactic medium was
reionized. The role of polarization will be discussed further in Section
\ref{sec:4}.

We follow the notation of BET97 and assume that a cosmological model is
specified by a set of parameters ${\bf s}$ . Let $P({\bf s} \vert prior)$
be the prior probability distribution of the parameters ${\bf s}$ and
$\cal L ({\bf s})$ the likelihood function defined by the experiment.  The
probability distribution of the parameters ${\bf s}$, taking into account
the new experimental information, is given by Bayes theorem $P({\bf s})
\propto {\cal L}({\bf s}) P({\bf s} \vert prior)$. If the errors $\Delta
{\bf s} \equiv {\bf s} - {\bf s}_0$ about the target model parameters
${\bf s}_0$ are small, an expansion of ${\rm ln} {\cal L}$ to quadratic
order about the maximum leads to the expression,
\begin{equation}
 {\cal L} \approx {\cal L}_m \exp\left[-{1 \over 2} \label{eq:3}
\sum_{ij}{F}_{ij}
\delta s_i \delta s_j\right]
\end{equation}
where $F_{ij}$ is the Fisher matrix,  given by derivatives of 
the CMB power spectrum with respect to the parameters ${\bf s}$ 
\begin{equation}
{F}_{ij} 
  = \sum_\ell { 1\over (\Delta {\rm C}_{\ell})^{2}} 
 {\partial {\rm C}_{\ell} \over \partial s_i} 
  {\partial {\cal C}_{\ell} \over \partial s_j}. \label{fish1}
\end{equation}
The quantity $\Delta C_\ell$ in equation~\ref{fish1}
is the standard error
on the estimate of $C_\ell$. For an experiment with $N$ frequency channels
(denoted by subscript $c$), angular resolution $(\theta_c)_{fwhm}$ and
sensitivity $(\sigma_c)_{pix}$ per resolution element 
($(\theta_c)_{fwhm} \times (\theta_c)_{fwhm}$ pixel) sampling a 
fraction $f_{sky}$,
\beglet 
\begin{eqnarray}
&&
 (\Delta {\rm C}_{\ell})^2 \approx 
{2\over (2\ell +1)f_{sky} }\, 
\left( {\rm C}_{\ell} 
+ \overline{w}^{-1} \overline{{\cal B}}_\ell^{-2}\right)^2, \label{eq:5a} 
\end{eqnarray}
%   \qquad \qquad \qquad \qquad \qquad\qquad\quad(5a) \\
\begin{eqnarray}
\overline{w} \equiv  \sum_c w_c, \quad \overline{{\cal B}}_\ell^2 
\equiv  \sum_{c} {\cal B}^2_{c\ell}w_c/\overline{w} ,   \label{eq:5b} \\
w_c \equiv (\sigma_{c,pix}\theta_{c,pix})^{-2}, \quad {\cal B}^2_{c\ell}
\approx e^{-\ell(\ell+1) /\ell_s^2}, \label{eq:5c}
\end{eqnarray}  
\endlet
(see Knox 1995 and BET97), where we have assumed that the experimental
beam profile ${\cal B}_c$ is Gaussian with width $\ell_s \equiv
\sqrt{8 {\rm ln} 2} (\theta_c)_{fwhm}^{-1}$. If we assume a uniform
prior, as we will do throughout this paper, then the covariance matrix
${\bf M} \equiv \langle \delta s \delta s^\dagger \rangle$ is the inverse
of the Fisher matrix ${\bf F}$. The standard deviation of a parameter
$s_i$, marginalized over uncertainties in the other parameters, is 
given by $\sigma_i = M_{ii}^{1/2}$.

Table \ref{tab1} lists the resolutions, $\theta_{fwhm}$, sensitivities
$\sigma_{pix}$ and the noise power parameters $w^{-1}$ for the various
frequency channels that we have adopted. The parameter set labelled
OMAP (`original MAP') is for the original specifications of the MAP
satellite, as given in the proposal of Bennet {\it et al}.  Parameters
for the current MAP design are listed under the heading CMAP (`current
MAP'). These include a significant improvement in the angular
resolution at all frequencies compared to the original design
specifications. We have listed the parameters for the $40$, $60$ and
$90$ GHz channels; the angular resolutions of the lower frequency MAP
channels are so much poorer ($0.93^\circ$ at $22$GHz and $0.68^\circ$
at $30$GHz) that they carry very little weight in the estimation of
cosmological parameters and so are ignored here. The main purpose of
these lower frequency channels is to monitor the free-free and
synchrotron contributions from the Galaxy.

%\begin{figure*}
\begin{table*}\label{tab1}
\bigskip
\centerline{\bf Table 1: Experimental Parameters}
\begin{center}
\begin{tabular}{|l|ccc|ccc|cccc|} \hline
 &  \multicolumn{3}{c|}{OMAP} & \multicolumn{3}{c|}{CMAP}&
 \multicolumn{4}{c|}{PLANCK}  \\ \hline
$\nu$ (GHz)   & $40$ & $60$  & $90$ &    $40$  &  $60$  & $90$ & 
$100$  &  $150$  & $220$ & $350$ \\
$\theta_{fwhm}$  & $32^\prime$ & $23^\prime$ & $17.4^\prime$ &  
 $28^\prime$  &  $21^\prime$  & $12.6^\prime$   &  $10.7^\prime$ & 
$8^\prime$  & $5.5^\prime$ & $5^\prime$ \\  
$\sigma_{pix}/10^{-6}$  & $7.6$ & $10.8$  & $16.3$ &    $6.6$  &  $12.1$  & 
$25.5$   &   $1.7$ & $2.0$ & $4.3$ & $14.4$ \\ 
$w_c^{-1}/10^{-15}$  & $4.9$  & $5.4$ & $6.8$ &   $2.9$ &  $5.4$ & $6.8$  & 
$0.028$ & $0.022$   & $0.047$ &  $0.44$  \\ \hline
\end{tabular}

\end{center}
\end{table*}
%\end{figure*}

For Planck, we have adopted parameters for the four lowest frequency
channels for the high-frequency instrument (HFI), as in the
current design submitted to ESA (Puget \etal 1998).  
The Planck payload is in a state of active
development and may change as the design of the Planck instruments
is refined. However, the  parameters listed
in Table \ref{tab1} give an accurate indication of what Planck is intended to
achieve. The major difference in the current Planck payload compared to that
described in the Phase A report is the addition of 
polarization sensitivity in the HFI instrument at $143$, 
$217$ and $545$ GHz and a much improved
performance of the low-frequency instrument (LFI), which covers the
frequency range $31$ -- $100$ GHz. The proposal to cool the LFI
to  $20$K  is expected to improve its sensitivity to within a factor
of two (or better) of that of the HFI. The
parameters listed in Table \ref{tab1} for Planck are thus likely to
underestimate the total sensitivity achievable by summing over all
frequency channels.

Evidently, there are some uncertainties in the experimental
parameters, particularly for Planck. We emphasise, therefore, that the
numbers given in this paper are meant to be illustrative of the
degeneracies among cosmological parameters resulting from experiments
with the characteristics of MAP and Planck, rather than a precise
analysis of these satellites.  A detailed analysis should include
systematic errors arising from Galactic and extra-Galactic
foregrounds, scanning strategy and map reconstruction, and
power-spectrum estimation.  A considerable amount of work along these
lines has been done already ({\it e.g.}  Bouchet, Gispert and Puget
1995, Tegmark and Efstathiou 1996, Bouchet \etal 1997, Hobson \etal
1998, Delabrouille, 1998) and will be developed in the future using
more precise models of the satellites.  In this paper, we assume that
the CMB power spectrum can be measured to the theoretical variance
given by equation (\ref{eq:5a}) over half of the sky, $f_{sky} =
0.5$. We therefore ignore any non-primordial contributions to the CMB
anisotropies (except in Section \ref{sec:5}), assuming that these can
be subtracted to high accuracy using the frequency information
available with both satellites.

The three sets of parameters in Table \ref{tab1} are also useful for
presentational purposes, since they help illustrate some of the key
points of this paper. The OMAP specifications are close to a threshold
at which there are large degeneracies between estimates of
cosmological parameters. These specifications have been retained for
their pedagogical value, even though the actual MAP performance is
expected to be much better.  As we show in Sections \ref{sec:3} and
\ref{sec:4}, the CMAP parameters (primarily the higher angular
resolution) lead to a significant improvement in cosmological
parameter estimates. Of course, this is the main motivation for
improving the angular resolution of MAP compared to the original
design specification.  The parameters for Planck define an experiment
that is almost `cosmic-variance limited', {\it i.e.} where random
experimental errors are less important than the sampling errors
arising from the fact that we can observe only one realisation of
microwave sky. In summary, we will see in later sections that
degeneracies among cosmological parameters are severe for OMAP, the
results for Planck are close to the best that can be achieved from
observations of the primary CMB anisotropies alone, and the results
for CMAP are intermediate between these two cases.

\section{Geometrical Degeneracy} \label{sec:3}

\subsection{Cosmological parameters} \label{sec:3.1}

We specify a spatially flat target cold dark matter
(CDM) model defined by the following 
parameters:

\noindent
$\bullet$ Spectral indices $n_s$ and $n_t$ for the scalar and tensor
components. The target model has $n_s = 1$ and $n_t= 0$.

\noindent
$\bullet$ The overall amplitude of the
CMB anisotropy spectrum defined by the average band
power $\langle \ell (\ell + 1) C_\ell/(2\pi ) \rangle^{1/2}$
over the range of multipoles $\ell_{max}$ accessible
to the experiment. We denote this amplitude, relative
to a COBE normalized model, by $Q$. The target model
has $Q=1$.

\noindent
$\bullet$ The ratio $r$ of the tensor and scalar components $r =
C_2^t/C_2^s$. The target model has $r = 0.2$.\footnote {In
inflationary models with an exactly uniform rate of acceleration the
ratio $r$ and the tensor and scalar spectral indices are related via
$r \approx - 7n_t = 7(1 - n_s)$ (see {\it e.g.} Davis \etal 1992,
Liddle and Lyth 1992, Bond \etal 1994). The parameters of the target
model are slightly inconsistent with these relations, though this is
of no consequence for any of the subsequent results in this
paper. }

\noindent
$\bullet$ The cosmological densities, $\Omega_i$, of the various
constituents of the universe, relative to the critical density of the
Einstein -de Sitter model. We include the contribution from baryons
$\Omega_b$, cold dark matter $\Omega_c$, a cosmological constant
$\Omega_\Lambda = \Lambda/(3H_0^2)$ (where $H_0$ is Hubble's
constant), and a density parameter that characterises the spatial
curvature of the Universe:
\begin{eqnarray}
&&  \Omega_k  =  1 - \Omega_b - \Omega_c - \Omega_\Lambda \, .\label{eq:6}
\end{eqnarray}
The CMB anisotropies are determined directly by {\it physical} densities,
$\omega_i$, which are related to the density parameters $\Omega_i$
by $\omega_i = \Omega_i h^2$, where $h$ is Hubble's constant
$H_0$ in units of $100\;{\rm km}{\rm s}^{-1}{\rm Mpc}^{-1}$. We
therefore use the physical densities $\omega_i$ to specify the target
model. The Hubble constant then becomes a secondary parameter that
is fixed by the constraint equation
\begin{eqnarray}
&& h^2  =  \omega_b  + \omega_c + \omega_k + \omega_\Lambda \, .\label{eq:7}
\end{eqnarray}
The target model has $\omega_b = 0.0125$, $\omega_c = 0.2375$, $\omega_k = 0$, 
$\omega_\Lambda = 0$, {\it i.e.} $\Omega_b = 0.05$, $\Omega_c = 0.95$,
$\Omega_k = 0$, $\Omega_\Lambda = 0$ and $h = 0.5$. The advantages
of using the densities $\omega_i$ for parameter estimation have been
discussed previously by BET97, and will become obvious in the next
two sections.

The target model is therefore specified by $8$ parameters, $4$
parameters defining the spectrum of fluctuations (two amplitudes and
two spectral indices); and $4$ density parameters specifying the
matter content and geometry of the universe. This is a relatively
small parameter set, but has been chosen to illustrate some
key points concerning parameter degeneracies. There would be little
value in including, for example, additional matter components such as
hot dark matter, or relativistic matter, since this would complicate
the discussion of degeneracies without adding anything substantially
new.

We have not included any parameters to specify reionization of the
intergalactic medium, {\it e.g.} an optical depth to Thomson
scattering, $\tau_{opt}$. This is because in the absence of
polarization measurements, the effects of reionization are almost
completely degenerate with a change in the amplitude of the
fluctuations ({\it i.e.} the product $Qe^{\tau_{opt}}$ is equal to a
constant). With polarization information, it is possible to break the
degeneracy between $Q$ and $\tau_{opt}$, provided $\tau_{opt}$ is
large enough (see ZSS97). The measurement of polarization also offers
a possibility of distinguishing between tensor and scalar modes
(Crittenden and Turok 1995, Seljak 1997a, Kamionkowski and Kosowsky
1998). If the polarization pattern can be reliably decomposed into
electric and magnetic components, it is possible to determine the
relative amplitudes of the scalar and tensor modes to much higher
accuracy than is possible from observations of the total CMB
anisotropy alone. Such a measurement would have a large qualitative
impact on the parameter estimates described in the next two sections,
significantly reducing the degeneracies between the tensor parameters
and other cosmological parameters. However, the tensor contribution to
the polarization signal is very small and concentrated to low $\ell$
values, so may be difficult to measure
against contaminating polarization signals of Galactic origin.

\subsection{The angular diameter distance degeneracy}\label{sec:3.2}

Two physical scales play a crucial role in determining the shape of
the CMB anisotropy spectrum. These are the sound horizon $r_s (a_r)$
at the time of recombination, and the angular diameter distance $d_A$
to the last scattering surface. (The quantity $a_r$ is the
cosmological scale factor at recombination, normalized to unity at the
present epoch). The sound horizon defines the physical scales for the
Doppler peak structure  
%(see {\it e.g} Hu and Sugiyama 1995)
that depends on the physical matter densities $\omega_b$ and $\omega_c$,
but not on the value of the cosmological constant or spatial curvature
since these are dynamically negligible at the time of recombination.  The
angular diameter distance depends on the net matter density, $\omega_m
\equiv \omega_b + \omega_c$, $\omega_k$ and $\omega_\Lambda$ and maps the
physical positions of the Doppler peaks to peaks in the angular power
spectrum $C_\ell$ as a function of multipole $\ell$.  This mapping leads
to a nearly exact `geometrical' degeneracy among cosmological
parameters, as has been noted by BET97 and ZSS97. Models will have
almost exactly the same CMB anisotropy spectra at high multipoles if they
satisfy the following criteria:

\noindent
[1] they have identical matter densities
$\omega_b$ and $\omega_c$;

\noindent
[2]  identical  primordial fluctuation spectra;

\noindent
[3] identical values of the parameter\footnote{Throughout this
paper we restrict our analysis to models with negative spatial
curvature (positive $\Omega_k$). There is no fundamental reason
for imposing this restriction, since it is straightforward to generalize
expressions such as equation (\ref{eq:8a}) to models with positive curvature.
However, all of the numerical examples described in this paper
have been computed with an updated version of the CMB anisotropy
code of Seljak and Zaldarriaga (1996) which is restricted to models
with negative spatial curvature.}
\beglet 
\begin{eqnarray}
&& {\cal R} = {\omega_m^{1/2} \over \omega_k^{1/2}}\; {\rm sinh}
\left [ \omega_k^{1/2} y \right ] \label{eq:8a} 
\end{eqnarray}
where
\begin{eqnarray}
&& y  = \int_{a_r}^{1} {da \over [\omega_m a + \omega_ka^2 + 
\omega_\Lambda a^4 + \omega_Q a^{1-3w_Q}]^{1/2}}. \label{eq:8b}
\end{eqnarray}
\endlet

\begin{figure*}

\vskip 2.8 truein

\includegraphics{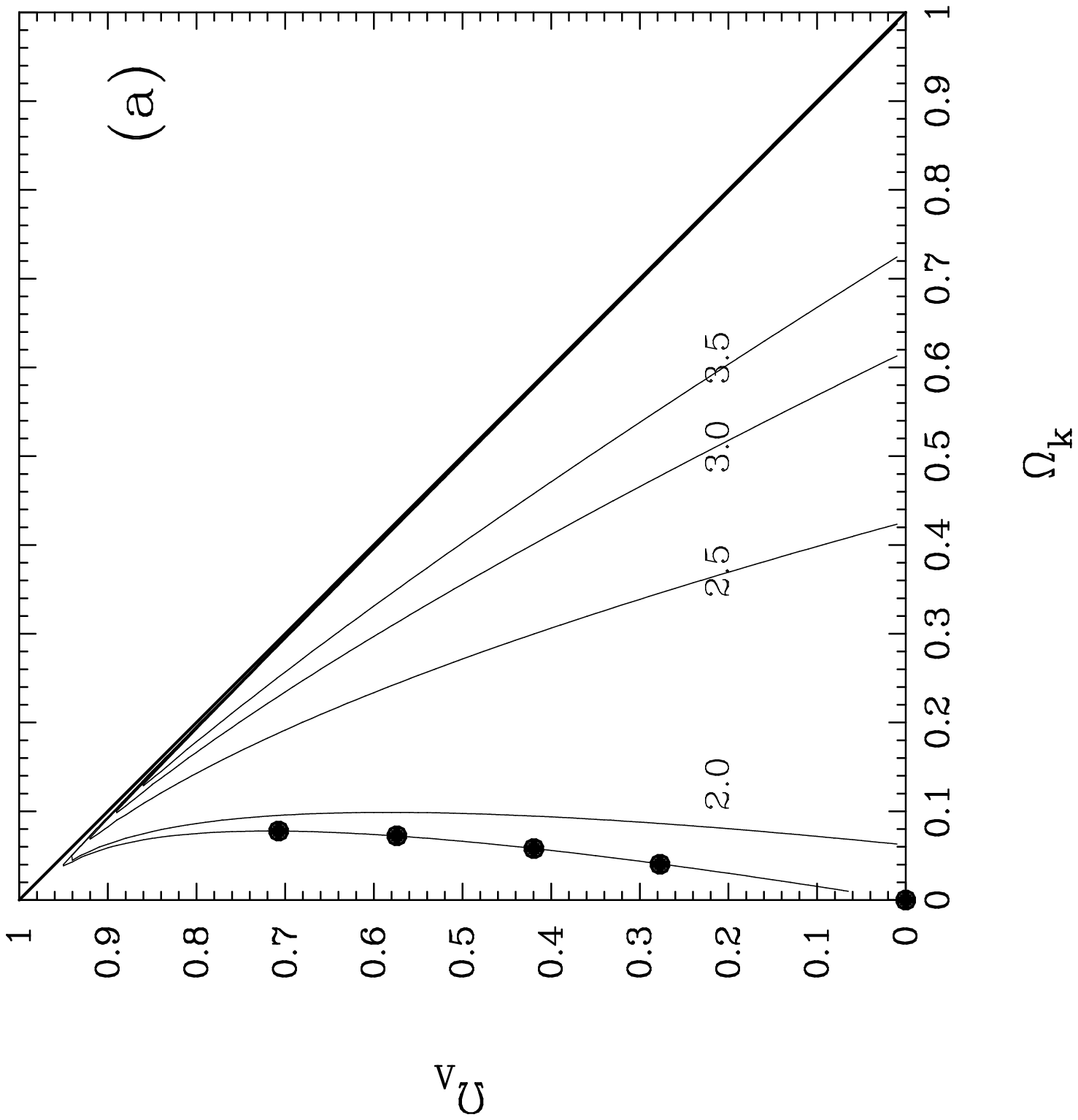}
\includegraphics{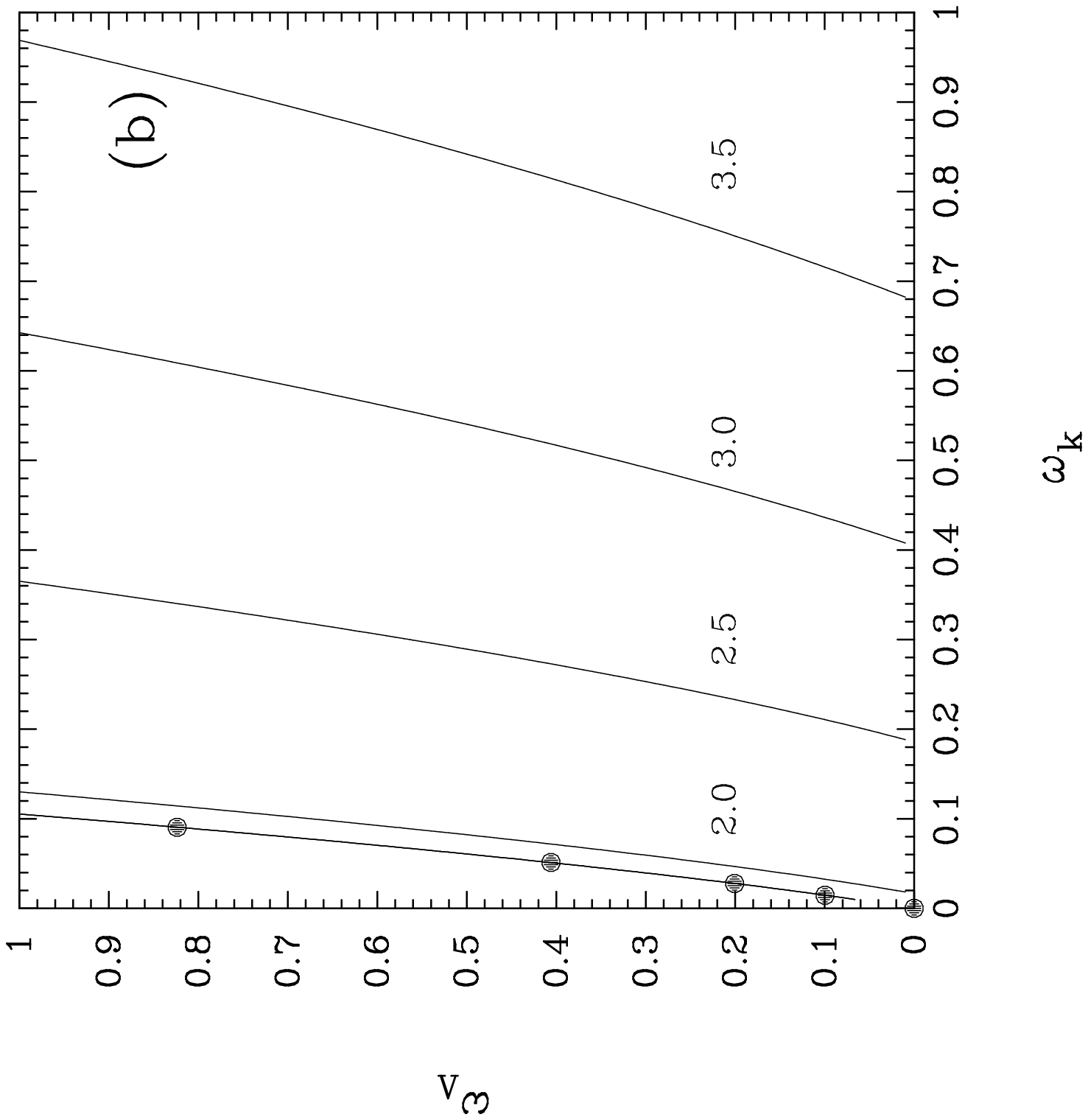}

\caption
{Figure \ref{figure1}a shows degeneracy 
lines of constant ${\cal R}$ in
the $\Omega_\Lambda$-$\Omega_k$ plane. The value of ${\cal R}$ is
given next to each line. In computing this figure, the matter 
density parameter is fixed by the constraint $\Omega_m = 1 - 
\Omega_k - \Omega_\Lambda$. Figure \ref{figure1}b shows lines of constant
${\cal R}$ in the $\omega_\Lambda$-$\omega_k$ plane for a 
universe with $\omega_m = 0.250$. The five dots in each figure
show the locations of the five models with nearly degenerate 
$C_\ell$ spectra plotted in Figure~\ref{figure2}. The target model with the 
parameters given in Section \ref{sec:3.1} is located at the origin in 
each panel.}
\label{figure1}
\end{figure*}

\noindent
Here $\omega_Q$ refers to a component of the energy density that obeys 
an equation of state with fixed pressure to density ratio,
$w_Q=p/\rho$, as might arise with scalar fields which are important at
late times in the universe, {\it e.g.}, Ratra and Peebles 1988 and
Caldwell, Dave and Steinhardt 1998 (who have dubbed such phenomena
quintessence). Note that $w_Q =-1/3$ gives the same contribution as
$\omega_k$ gives in $y$, but the geometrical aspects of the $\omega_k$
angle-distance relation are absent, yielding a behaviour more similar
to that of $\omega_\Lambda$.  Since it therefore does not add
qualitatively new features to the discussion we ignore $\omega_Q$
terms in what follows.  Equations (\ref{eq:8a},\ref{eq:8b}) are
special cases of more general expressions for the locations of the
Doppler peaks given in Section \ref{sec:4}.  Provided the above
conditions are satisfied, the only differences between the CMB power
spectra of models with different values of $\Omega_\Lambda$ and
$\Omega_k$ will be those at low multipoles $\ell \simlt 30$ arising from 
the late-time Sachs-Wolfe effect (see Bond 1996, Hu, Sugiyama and
Silk 1997, and references therein).

\begin{figure*}

\vskip 4.4 truein

\includegraphics{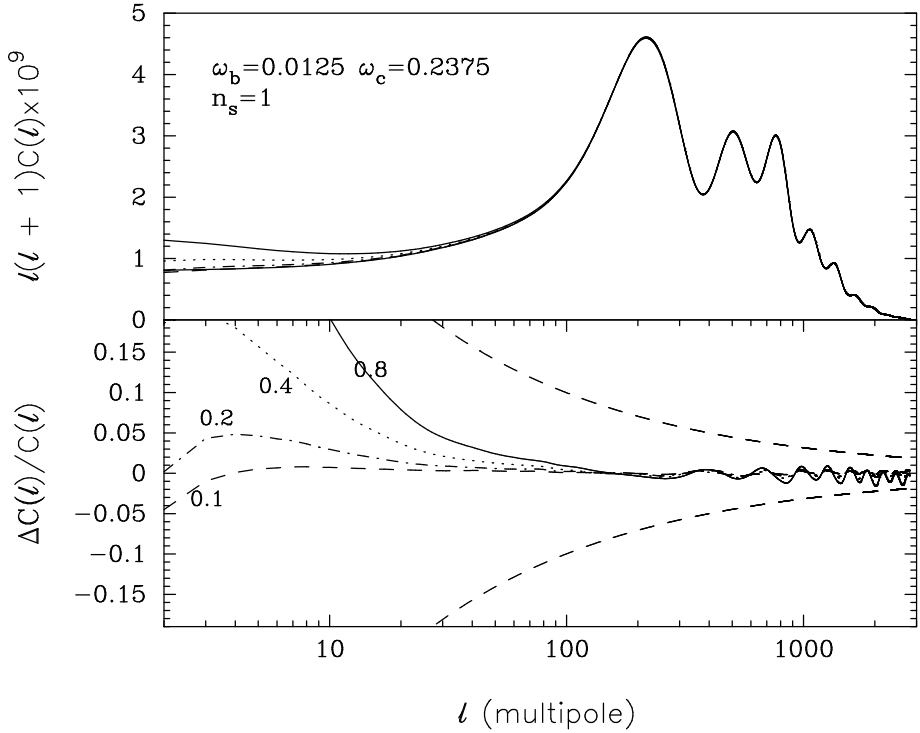}

\caption{The upper panel shows the scalar power spectrum for the
nearly degenerate models shown by the dots in Figures
\ref{figure1}. The target model has $\omega_\Lambda = \omega_k = 0$,
$\omega_b = 0.0125$ and $\omega_c = 0.2375$ and $n_s = 1$. The lower
panel shows the residuals of these open universe models with respect
to the spatially flat target model. The numbers next to each curve
give the value of $\omega_\Lambda$ for each model. The two lines with
long dashes show the standard deviation of the residuals from cosmic
variance alone.}
\label{figure2}
\end{figure*}

This is illustrated in Figures~\ref{figure1} and \ref{figure2}. Figure 1 shows lines of
constant ${\cal R}$ in the $\Omega_\Lambda$ - $\Omega_k$ plane (Figure
\ref{figure1}a) and in the $\omega_\Lambda$ - $\omega_k$ plane (Figure \ref{figure1}b), for
a fixed value of $\omega_m = 0.25$. Figure \ref{figure1}a is the more useful,
because the degeneracy lines are fixed by the two parameters
$\Omega_\Lambda$ and $\Omega_k$, with the matter density specified by
the constraint $\Omega_m = 1 - \Omega_k -
\Omega_\Lambda$. The degeneracy lines in the space defined
by the physical densities $\omega_m$, $\omega_k$ and 
$\omega_\Lambda$ require specification of all three densities
or, alternatively, two densities and a value of the Hubble constant
(see equation \ref{eq:7}). We have therefore  assumed the value of $\omega_m$ 
of our target model in computing Figure \ref{figure1}b.

The five dots in each panel of Figure \ref{figure1} show the locations of models
with nearly degenerate $C_\ell$ spectra. These models have the same
matter densities as the target model, $\omega_b = 0.0125$ and
$\omega_c =0.2375$, and identical scalar spectral index $n_s = 1$ (we
compute only the scalar component for this comparison). The five
models have different values of $\omega_k$ and $\omega_\Lambda$ chosen
to lie along the degeneracy line of ${\cal R} = 1.94$ of the target
model, as shown in the figures.  The resulting CMB power spectra are
plotted in Figure \ref{figure2}.

As expected, the power spectra plotted in Figure \ref{figure2} are
indistinguishable except at small multipoles, $\ell \simlt 30$. The
lower panel compares the residuals $\Delta C_\ell/C_\ell$ for these
models compared to the spatially flat target model. The two unlabelled
dashed lines show the error on the residuals arising from cosmic
variance alone, $\Delta C_\ell/C_\ell = \pm(2 /(2 \ell +1
))^{1/2}$. There are several points worth noting from this figure:

\noindent
[1] The differences between the models at high multipoles ($\ell
\simgt 200$) are small and arise from numerical errors. These include
small errors in the computation of the degeneracy parameter ${\cal
R}$, but arise predominantly from numerical errors in the CMBFAST
code.

\noindent
[2] Although the numerical errors at high multipoles are much smaller
than the cosmic variance errors, they accumulate in the computation of
the Fisher matrix (equation~\ref{fish1}) and can erroneously break the
geometrical degeneracy. Thus, numerical errors in what should be an
exact degeneracy can lead to extremely misleading results, in which
the errors in the density parameters $\omega_\Lambda$ and $\omega_k$
appear weakly correlated and extremely small. This  problem is
particularly acute for a Planck-type experiment in which the $C_\ell$
estimates are nearly cosmic variance limited to high multipoles.

\noindent
[3] The differences in the power spectra at low multipoles, $\ell
\simlt 30$, arise from a real physical effect -- the late time
Sachs-Wolfe effect -- and are not caused by numerical errors. However,
the differences are smaller than the cosmic variance limits and so it
is impossible to differentiate between these models at a high
significance level from measurements of the CMB anisotropies
alone. (This point is quantified in detail in Figure \ref{figure3}
below).

\noindent
[4] The power spectra for the low density cosmological models
assume a particular model of the origin of the primordial fluctuation 
spectrum (in this case, power-law fluctuations in the potentials $\Phi$ and
$\Psi$). If we were to allow complete freedom
to adjust the shape of the power spectrum on large spatial scales, then
we could compensate for the differences caused by the late-time Sachs
Wolfe effect. The existence of the late-time Sachs-Wolfe effect cannot
be used to distinguish between different cosmological models unless specific
assumptions are made concerning the shape of the fluctuation spectrum.

The futility of breaking the geometrical degeneracy from linear CMB
anisotropies is illustrated in Figure \ref{figure3}. In this figure,
we have computed a grid of scalar $C_\ell$ spectra varying the
parameters $\omega_\Lambda$ and $\omega_k$, whilst fixing the
remaining parameters, $\omega_b$, $\omega_c$ and $n_s$ to those of the
spatially flat target model. Figure \ref{figure3} shows the exact
likelihood ratio,
\begin{eqnarray}
 - 2 {\rm ln } \left ( {{\cal L} \over {\cal L}_{max}} \right )
  &= &  \sum_{\ell \le \ell_{max}} { (C_\ell({\bf s}) - C_\ell 
({\bf s}_0))^2 
\over (\Delta C_\ell)^2 }, \label{eq:9}
\end{eqnarray}
for an experiment with the OMAP parameters listed in Table \ref{tab1}
and $\ell_{max}$ set to $1000$. The contours show approximate $1$, $2$
and $3\sigma$ contours in the $\omega_\Lambda$ and $\omega_k$ plane
assuming all other parameters are known ({\it i.e.} we show contours
where $-2{\rm ln}({\cal L}/{\cal L}_{max})$ equals
$2.3$, $6.2$ and $11.8$ corresponding to $68\%$, $95\%$ and $99.7\%$
confidence regions if $-2{\rm ln}({\cal L}/{\cal L}_{max})$ is
$\chi^2$ distributed with two degrees of freedom).
The $1\sigma$ contours are
broken up artificially by the discrete size of the grid over which the
$C_\ell$'s were computed numerically.  The $2\sigma$ contour limits
$\omega_\Lambda$ to be less than about $0.7$ and $\omega_k$ to be less
that about $0.1$. The contours follow the line of constant ${\cal R}$
computed in Figure \ref{figure1}b and show that measurements of the
CMB anisotropies alone cannot remove the geometrical degeneracy
between $\omega_\Lambda$ and $\omega_k$. Improving the accuracy of the
CMB measurements simply causes the likelihood contours to narrow
around the degeneracy lines. Although this has important consequences
for the estimation of other cosmological parameters, improving the
accuracy of the experiment does not help to remove the geometrical
degeneracy.  For all practical purposes, the geometrical degeneracy in
linear theory is exact and can be removed only by invoking other
constraints on the geometry of the Universe ({\it e.g.} as derived from 
Type Ia supernovae, Perlmutter \etal 1997, 1998, see Section
\ref{sec:6}), or from the effects of gravitational lensing on the CMB
(Stompor and Efstathiou 1998, Metcalf and Silk 1998).

\begin{figure}

\vskip 3.2 truein

\includegraphics{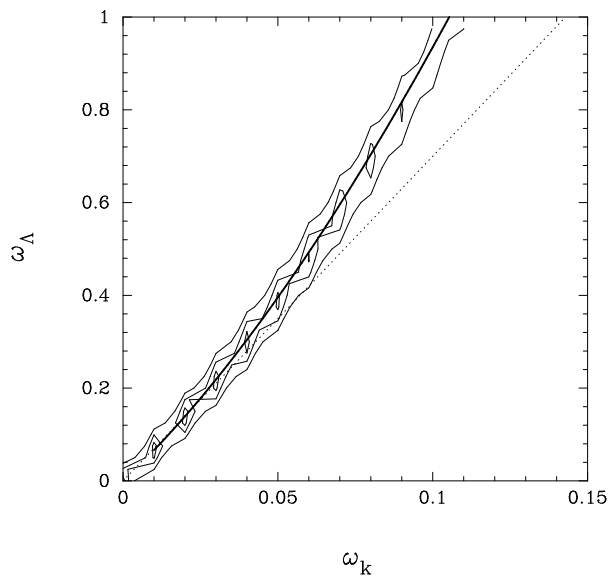}

\caption{Likelihood ratio contours in the $\omega_\Lambda$-$\omega_k$
plane for models containing only scalar modes. The models have 
fixed values of $\omega_b=0.0125$ and $\omega_c=0.2375$. The likelihood
ratio for each model has been computed assuming the experimental parameters
of OMAP (Table \ref{tab1}) and our standard spatially flat target
model. The contours show where $-2{\rm ln}({\cal L}/{\cal L}_{max})$ has
values of $2.3$, $6.2$ and $11.8$ corresponding approximately to
$1\sigma$, $2 \sigma$ and $3\sigma$ contours assuming all other parameters
are known.  The full line shows the degeneracy curve with the same value
of ${\cal R}$ as the target model ({\it cf} Figure \ref{figure1}b) and the dashed line
shows the Taylor series approximation to this curve, $\omega_\Lambda = 7
\omega_k$ (see Section \ref{sec:3.3}).}
\label{figure3}
\end{figure}

\subsection{Parameter estimation including spatial curvature}\label{sec:3.3}

In this Section, we describe how the Fisher matrix formalism can be
adapted to handle the geometrical degeneracy. This can be achieved by
redefining the variables ${\bf s}$ so that one of the derivatives
$\partial C_\ell/\partial s_i$ is computed along the direction of the
degeneracy ({\it i.e.} along lines of constant ${\cal R}$, as in the
examples plotted in Figure \ref{figure2}).  The derivative of $C_\ell$ along the
degeneracy line can then be set to zero at high multipoles by
construction, eliminating the effects of numerical errors in computing
the elements of the Fisher matrix. This technique has been used
previously by us in BET97, and gives extremely stable results.

To illustrate the method, we analyse the errors in the cosmological
parameters for small variations around the target model defined in
Section \ref{sec:3.1}. The target model has $\Omega_k = 0$, $\Omega_\Lambda = 0$ 
and  $\Omega_m = 1$, and so the derivatives of ${\cal R}$ are
\beglet 
\begin{eqnarray}
&& \left 
( {\partial {\cal R} \over \partial \Omega_k} \right )_t = 1\, , \quad
\left ( {\partial {\cal R} \over \partial \Omega_\Lambda} \right )_t = 
- {1 \over 7} \label{eq:10a}
\end{eqnarray}
where the subscript $t$ denotes that the derivatives are computed
at the  parameter values of the target model. For small variations
around the target model, the degeneracy lines in ${\cal R}$ follow 
\begin{eqnarray}
&& \delta \Omega_k = {1 \over 7} \delta \Omega_\Lambda,
\qquad  \delta \omega_k =   {1 \over 7} \delta \omega_\Lambda.
\label{eq:10b} 
\end{eqnarray}
\endlet
It is therefore useful to define an auxiliary density
\begin{eqnarray}
&& \omega_D  =  7 \omega_k - \omega_\Lambda \,  \label{eq:11}
\end{eqnarray}
to replace $\omega_k$ in the Fisher matrix analysis (where the
subscript, D, denotes `degeneracy') . For small
variations around the target model, the constraint $\omega_D=0$
defines the degeneracy line ${\cal R}={\rm constant}$. The CMB spectra
plotted in Figure \ref{figure2} all have $\omega_D \approx 0$ and identical values
of $\omega_b$ and $\omega_c$. They can therefore be used to define the
derivative $\partial C_\ell/\partial \omega_\Lambda$, which can be set
to zero at, for example, $\ell \simgt 200$ to suppress the effects of
numerical errors. The results of such an analysis for the experiments
summarized in Table \ref{tab1} are listed in Table \ref{tab2}.

%\bigskip
\begin{table}\label{tab2}
\centerline{\bf Table 2: $1\sigma$ errors on estimates of }
\centerline{\bf cosmological parameters}
\begin{center}
\begin{tabular}{|cccc|} \hline
Parameter  &  OMAP & CMAP & PLANCK   \\ \hline
$\delta \omega_b/\omega_b$        & $0.075$      & $0.042$       
& $0.0064$          \\
$\delta \omega_c/\omega_c$        & $0.24\;\;$   & $0.13\;\;$    
& $0.020\;\;$       \\
$\delta \omega_D$                 & $0.17\;\;$   & $0.097$       
& $0.014\;\;$       \\
$\delta \omega_\Lambda$           & $1.23\;\;$   & $1.20\;\;$    
& $1.09\;\;\;$        \\
$\delta Q$                        & $0.013$      & $\;\;0.0041$      
& $\;0.0013$          \\
$\delta r$                        & $0.41\;\;$       & $0.27\;\;$        
& $0.15\;\;$        \\
$\delta n_s$                      & $0.059$      & $0.034$       
& $\;0.0065$           \\
$\delta n_t$                      & $1.015$      & $0.998$       & $0.911$    
\\ \hline
\end{tabular}
\end{center}
\end{table}

The results from Table \ref{tab2} show that the $1\sigma$ errors in
$\omega_\Lambda$ are close to unity and extremely insensitive to the
experimental details.  This is in agreement with the likelihood
analysis of Figure \ref{figure3}. The large error in $\omega_\Lambda$
thus reflects the geometrical degeneracy. In contrast, the error on
the auxiliary density parameter $\omega_D$ decreases from $0.17$ for
the OMAP experimental parameters, to $0.014$ for the Planck
parameters. The error in $\omega_D$ is set by the accuracy to which
the parameter ${\cal R}$ can be determined by the experiment, {\it
i.e.} by the ability of the experiment to fix the positions of the
Doppler peaks.  An experiment such as Planck samples the entire
Doppler peak structure of $C_\ell$ and so can determine ${\cal R}$
(and hence $\omega_D$) to extremely high accuracy.

\begin{figure}

\vskip 3.6 truein

\includegraphics{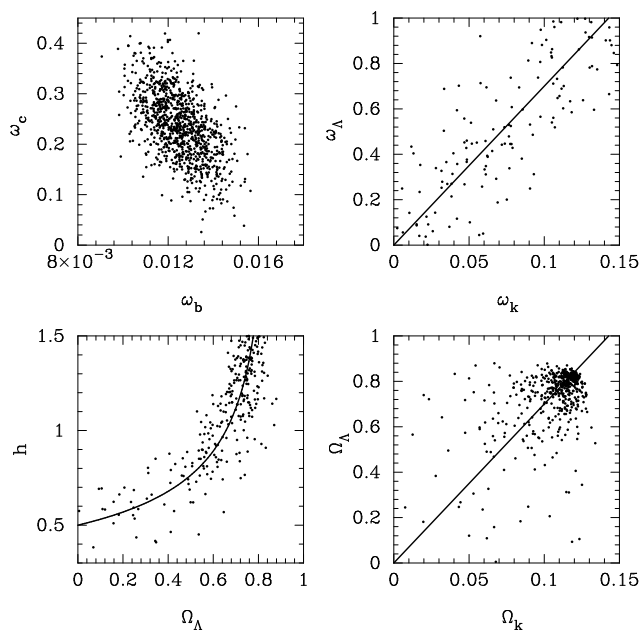}

\caption{ Results of Monte-Carlo
simulations of parameter estimation as described in the text assuming
the parameters of OMAP as given in Table \ref{tab1}.  The panels show
correlations between various pairs of parameters. The lines in the
$\omega_\Lambda$-$\omega_k$ and $\Omega_\Lambda$-$\Omega_k$ plane show
the degeneracy lines given by equation (\ref{eq:10b}). The line in the
$h$-$\Omega_\Lambda$ plane shows the degeneracy line given by equation~(\ref{eq:12}).}
\label{figure4}
\end{figure}

\begin{figure*}

\vskip 3.5 truein

\includegraphics{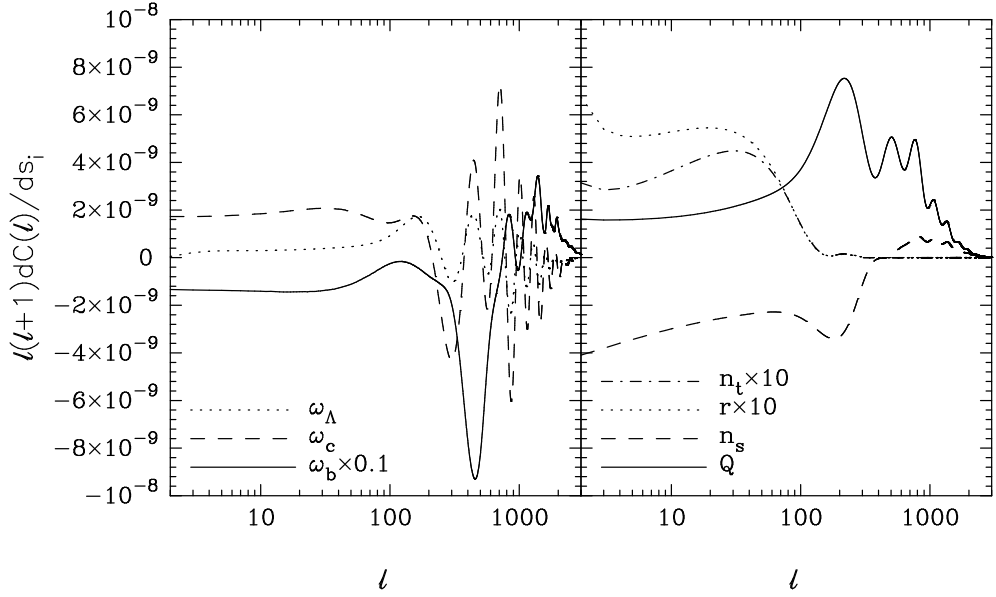}

\caption{Derivatives of $C_\ell$ with respect to the seven
parameters of the spatially flat target model defined in the text.
The derivatives with respect to $\omega_b$, $n_t$ and $r$ have been
multiplied by the factor indicated in the figure ({\it e.g.} the
derivative $\partial C_\ell/\partial \omega_b$ has been divided by 
a factor of ten).}
\label{figure5}
\end{figure*}

Another important consequence of the geometrical degeneracy is that it
sets a fundamental limitation on a determination of the Hubble
constant from measurements of the primary CMB anisotropies alone.
This arises because the Hubble constant is a secondary parameter,
determined from the constraint equation (\ref{eq:7}). The error in the
determination of $h$ will be dominated by the error in
$\omega_\Lambda$, which, as we have seen above, is essentially
unconstrained because of the geometrical degeneracy. The value of $h$
determined from the CMB anisotropies is therefore degenerate with the
value of $\Omega_\Lambda$. For satellite experiments such as MAP and
Planck, we can assume to a good approximation that the parameters
$\omega_b$, $\omega_c$ and $\omega_D$ are determined exactly. Equation
(\ref{eq:7}) then provides a relation between $h$ and $\Omega_\Lambda$
\begin{eqnarray}
&& h  =  {h_t / ( 1 - {8 \over 7} \Omega_\Lambda)^{1/2}} \, .\label{eq:12}
\end{eqnarray}
Here, $h_t$ denotes the Hubble constant of the target model. Formally, 
equation (\ref{eq:12}) is valid for small variations around the target model
parameters, but in practice it is reasonably accurate for 
large variations in $\Omega_\Lambda$ (because the degeneracy lines in
Figure \ref{figure1}b are very  nearly straight lines). Equation (\ref{eq:12}) is independent
of the details of the experiment. The scatter around the relationship
(12), however, is set by  the errors of the density parameters (mainly errors
in  $\omega_D$) and so depends on the experimental parameters.

These points are illustrated in Figure \ref{figure4} which shows the results of
parameter estimation applied to a set of simulated $C_\ell$ spectra.
Here we have generated $C_\ell$ spectra of our target model with
errors computed for the OMAP experiment and we have determined the
cosmological parameters for each model by $\chi^2$ minimisation,
assuming that the theoretical spectra of neighbouring points can be
approximated by the Taylor series expansion 
\begin{eqnarray}
&& C_\ell ({\bf s}_i)  =
C_\ell({\bf s}_0) + \left ( {\partial C_\ell \over 
\partial  s_i} \right )\cdot \Delta s_i. \label{eq:13}
\end{eqnarray}
This procedure is designed by construction to mimic the Fisher matrix
approach exactly.  Figure \ref{figure4} shows that $\omega_c$ and
$\omega_b$ are slightly anticorrelated, with errors as given in Table
\ref{tab2}. The parameters $\omega_\Lambda$ and $\omega_k$ are
extremely highly correlated along the degeneracy line defined by
equation (\ref{eq:10b}). The scatter about this line is determined by
the errors in $\omega_D$.  The parameters $h$ and $\Omega_\Lambda$ are
also strongly correlated, following equation (\ref{eq:12}) to high
accuracy. The final plot in Figure \ref{figure4} shows the parameters
$\Omega_\Lambda$ plotted against $\Omega_k$. This plot may seem a
little strange, because it shows a clump of points at $\Omega_\Lambda
\sim 0.8$ and $\Omega_k \sim 0.13$, well away from the target model
parameters $\Omega_\Lambda = \Omega_k = 0$.  This is because the
errors in $\omega_\Lambda$ and $\omega_D$ are symmetrical by
construction and we determine $\Omega_\Lambda$ and $\Omega_k$ from
$\omega_\Lambda$, $\omega_D$ and the Hubble constant as computed from
the constraint relation of equation (7). Since the error in
$\omega_\Lambda$ is large, the points in the $\Omega_\Lambda$--
$\Omega_k$ plane must, by construction, cluster around the values
$\Omega_\Lambda \approx 7/8$, $\Omega_k \approx 1/8$.

In summary, the geometrical degeneracy leads to a fundamental
indeterminacy in the parameters $\Omega_\Lambda$ and $\Omega_k$. For
spatially flat target models, this leads to a near indeterminacy of
$\Omega_\Lambda$ and an error on $\Omega_k$ of $\sim 0.1$ irrespective
of the accuracy of the CMB experiment. The indeterminacy of
$\Omega_\Lambda$ leads to an indeterminacy of the Hubble constant via
the constraint relation (7). As a corollary, independent constraints
on $H_0$ and $\Omega_\Lambda$ from observations other than the CMB
anisotropies can be used to break the geometrical degeneracy.

The discussion of this Section also shows that tables  designed to
demonstrate the power of various experiments, such as Table \ref{tab2}, must be
interpreted with caution. Evidently the errors on some
cosmological parameters are extremely sensitive to the assumed  parameter
set. For example, restricting to models with zero cosmological constant,
or  curvature, would have a very large  effect on the accuracy of the
determination of $H_0$.  Similar remarks apply to other parameters and to
near degeneracies other than the geometrical degeneracy described here.
This is the topic of the next Section.

\section{Eigenvectors of the Fisher Matrix and Parameter Degeneracies} \label{sec:4}

\subsection{Motivation}\label{sec:4.1}

The geometrical degeneracy described in the previous Section
represents an extreme case of a near exact degeneracy between
cosmological parameters. In this Section, we will investigate partial
degeneracies among other parameters showing, for example, how
uncertainties in the parameters defining the primordial spectra couple
to the cosmological densities in baryons and CDM.  The results of the
previous section show that the geometrical degeneracy is so nearly
exact that very little is lost in analysing degeneracies among other
parameters by restricting to a spatially flat background
universe. Such a restriction merely removes the geometrical degeneracy from 
consideration without affecting any other aspect of the analysis.
In this Section, we therefore analyse the parameter errors for the
same spatially flat target model as defined in Section \ref{sec:3.1}
but restricted to a set of $7$ parameters, $n_s$, $n_t$, $Q$, $r$,
$\omega_b$, $\omega_c$ and $\omega_\Lambda$. The derivatives of
$C_\ell$ with respect to these seven parameters are plotted in Figure
\ref{figure5}.

\begin{table*}\label{tab3}
\bigskip
\centerline{\bf Table 3: Principal components}
\begin{center}
\begin{tabular}{cccccccc} \hline \hline
\noalign{\medskip}
\multicolumn{8}{c}{OMAP errors} \\
\noalign{\medskip}   \hline 
\noalign{\medskip}
$1/\sqrt\lambda$ & $\omega_b$ & $\omega_c$ & $\omega_\Lambda$ &  $Q$      &  $n_s$     & $r$      & $n_t$ \\
2.27E-4   &  0.998$^*$ & -1.417E-2  & -1.103E-2        & -6.376E-2 &  4.919E-3  & -2.972E-4 & -2.599E-4   \\
3.14E-3   &  5.363E-2  & -0.284  &  3.051E-3        &  0.871$^*$ & -0.397  & 1.258E-2 &  1.188E-3   \\
4.46E-3   &  2.727E-2  &  $\;\;$0.886$^*$ &  0.321        &  1.513E-2 & -0.294  & 2.910E-2 &  2.633E-2   \\
1.91E-2   &  2.722E-2  &  $\;$0.135     &    0.214           & $\;$0.424    &  $\;$0.829$^*$ & -0.198    &  -0.167     \\
4.32E-2   & -6.283E-3  & -0.312     &  $\;$0.821$^*$       & -0.183    & -0.191     & 0.300    &  -0.261     \\
3.18E-1   &  3.315E-3  & -0.134     &  0.416           &  1.798E-2 &  0.177     & 0.565$^*$&   $\;$0.676$^*$ \\
1.06      &  6.271E-4  & -1.768E-2  &  5.901E-2        &  4.918E-3 &  2.753E-2  & 0.742$^*$&  -0.667$^*$ \\ \hline\hline
%\noalign{\medskip}
%\multicolumn{8}{c}{CMAP errors} \\
%\noalign{\medskip}   \hline 
%\noalign{\medskip}
%$1/\sqrt\lambda$ & $\omega_b$ & $\omega_c$ & $\omega_\Lambda$ &  $Q$       &  $n_s$     & $r$      & $n_t$ \\
%1.56E-4   &  0.998$^*$ & -1.722E-2  & -1.120E-2        & -6.180E-2  &  1.995E-3  & -1.396E-4  & -1.222E-4   \\
%2.73E-3   &  3.851E-2  & -0.549$^*$ & -0.119           &  $\;$0.788$^*$ & -0.248  &  5.511E-3  &  5.354E-3   \\
%3.09E-3   &  4.901E-2  &  $\;$0.780$^*$ &  $\;$0.279           &  $\;$0.517$^*$ & -0.209     &  1.588E-2  &  1.460E-2   \\
%8.64E-3   &  1.814E-2  &  9.132E-2  & -0.167           &  0.328     &  $\;\;$0.921$^*$ & -6.501E-2  & -5.721E-2   \\
%3.10E-2   &  4.719E-3  & -0.254     &  $\;\;$0.847$^*$       & -9.028E-4  &  $\;$0.137     & -0.340     & -0.290     \\
%1.86E-1   &  2.819E-3  & -0.133   &  $\;$0.403           &  1.209E-2  &  $\;$0.166     &  $\;$0.648$^*$ &  $\;\;$0.611$^*$ \\
%0.97      & -1.545E-4  &  5.530E-3  & -1.784E-2        & -9.055E-4  &
%-5.786E-3  & -0.679$^*$ &  $\;\;$0.734$^*$ \\ \hline \hline
\noalign{\medskip}
\multicolumn{8}{c}{CMAP errors} \\
\noalign{\medskip}   \hline 
\noalign{\medskip}
$1/\sqrt\lambda$ & $\omega_b$ & $\omega_c$ & $\omega_\Lambda$ &  $Q$       &  $n_s$     & $r$      & $n_t$ \\
1.89E-4   &  0.998$^*$ & -1.592E-2  & -1.113E-2        & -6.241E-2  &  3.257E-3  & -2.052E-4  & -1.795E-4   \\
2.98E-3   &  4.883E-2  & -0.366     & -3.678E-2        &  $\;$0.862$^*$ & -0.344  &  1.005E-2  &  9.545E-3   \\
3.82E-3   &  3.571E-2  &  $\;\;$0.870$^*$ &  $\;$0.314   &  0.279     & -0.252    &  2.238E-2  &  2.0370E-2   \\
1.22E-2   &  2.471E-2  &  0.149     & -6.142E-2        &  0.415     &  $\;\;$0.886$^*$ & -9.596E-2  & -8.279E-2   \\
3.39E-2   &  2.270E-3  & -0.261     &  $\;\;$0.849$^*$ & -4.378E-2  &  5.800E-2     & -0.345     & -0.295     \\
2.46E-1   &  3.025E-3  & -0.136     &  $\;$0.417       &  1.385E-2  &  $\;$0.171     & $\;\;$0.616$^*$ &  $\;\;$0.631$^*$ \\
0.997      & -2.923E-4  &  1.032E-2  & -3.328E-2       & -1.769E-3  & -1.260E-2  & -0.701$^*$ &  $\;\;$0.712$^*$ \\ \hline \hline
%\noalign{\medskip}
%\multicolumn{8}{c}{PLANCK  errors} \\ 
%\noalign{\medskip}  \hline
%\noalign{\medskip} 
%$1/\sqrt\lambda$ & $\omega_b$ & $\omega_c$ & $\omega_\Lambda$ &  $Q$       &  $n_s$     & $r$      & $n_t$ \\ 
%4.03E-5   &  0.999$^*$ & -3.129E-2  & -1.713E-2        &  3.203E-2     &   1.860E-2  & -1.061E-5  & -9.117E-6   \\
%5.23E-4   &  3.211E-2  &  0.958$^*$ &  0.274           &  6.587E-2     &   2.747E-2  &  2.778E-4  &  2.474E-4   \\
%7.39E-4   & -3.756E-2 &  -5.996E-2  & -4.042E-2        &  $\;\;$0.977$^*$ &  0.196     &  8.063E-4  &  7.467E-4   \\
%4.31E-3   & -1.150E-2  & -4.409E-2  &  0.105           & -0.194     &  $\;\;$0.973$^*$ & -3.372E-2  & -3.003E-2   \\
%5.17E-2   &  8.362E-3  & -0.274     &  $\;\;$0.955$^*$ &  4.425E-2  &  -0.106     & 5.011E-3  & 4.980E-3     \\
%5.53E-2   & -5.352E-4  & -3.551E-4  & -2.064E-3        & -1.018E-2  &  4.450E-2  &  $\;$0.760$^*$ &  0.648$^*$ \\
%0.75      & -1.424E-5  &  9.943E-5  & -4.133E-4        & -2.580E-4  &
%9.979E-4  & -0.649$^*$ &  0.761$^*$ \\ \hline \hline
\noalign{\medskip}
\multicolumn{8}{c}{PLANCK  errors} \\ 
\noalign{\medskip}  \hline
\noalign{\medskip} 
$1/\sqrt\lambda$ & $\omega_b$ & $\omega_c$ & $\omega_\Lambda$ &  $Q$       &  $n_s$     & $r$      & $n_t$ \\ 
4.09E-5   &  0.999$^*$ & -3.110E-2  & -1.707E-2        &  3.253E-2     &   1.847E-2  & -1.088E-5  & -9.348E-6   \\
5.29E-4   &  3.195E-2  &  0.958$^*$ &  0.274           &  6.588E-2     &   2.740E-2  &  2.852E-4  &  2.541E-4   \\
7.42E-4   & -3.704E-2 &  -5.976E-2  & -4.110E-2        &  $\;\;$0.977$^*$ &  0.196     &  8.106E-4  &  7.508E-4   \\
4.32E-3   & -1.172E-2  & -3.611E-2  &  7.679E-2           & -0.195     &  $\;\;$0.976$^*$ & -3.396-2  & -3.025E-2   \\
5.25E-2   &  7.997E-3  & -0.275     &  $\;\;$0.957$^*$ &  3.922E-2  &  -7.733E-2     & 3.716E-3  & 3.843E-3     \\
5.54E-2   & -5.331E-4  & -4.795E-4  & -2.665E-3        & -1.017E-2  &  4.461E-2  &  $\;$0.760$^*$ &  0.648$^*$ \\
0.75      & -1.424E-5  &  1.003E-4  & -4.164E-4        & -2.579E-4  &  9.970E-4  & -0.649$^*$ &  0.761$^*$ \\ \hline \hline
\end{tabular}
\caption{Components of the projection vector $U^\dagger$ relating the original variables $\hat s_i$ (with means subtracted,
$\hat s_i = s_i - \langle s_i \rangle$) to the principal components
${\bf X}$. The eigenvalues are listed in the first column. We have marked 
each component of $U^\dagger$ that is greater than $0.5$ with an asterisk
to show which physical variables contribute significantly to each 
principal component.}

\end{center}
\end{table*}

\subsection{Principal components}\label{sec:4.2}

The Fisher matrix $F$ is a symmetric $n \times n$ matrix and so can
be reduced to diagonal form,
%\footnote{There should be no reason to confuse
%the matrix of eigenvalues $\Lambda$ defined in this equation with the
%cosmological constant.}
\begin{eqnarray}
F  &= & U \Lambda U^\dagger, \qquad \Lambda = {\rm diag}(\lambda_1,
\lambda_2, ...., \lambda_n) \, , \label{eq:14}
\end{eqnarray}
where $U$ is the matrix in which the m'th row is the eigenvector ${\bf
u}_m$ corresponding to the eigenvalue $\lambda_m$. The eigenvectors
are assumed ordered so that $\lambda_1 \ge \lambda_2 \ge \lambda_3
... \ge \lambda_m$. From our original variables ${\bf s}$ we can
therefore construct a set of new variables ${\bf X}$ that are
orthogonal to each other,
\begin{eqnarray}
 {\bf X}  &= & U^\dagger{\bf s}, \qquad 
{\bf s}^\dagger F {\bf s}  = {\bf X}^\dagger \Lambda {\bf X}.\label{eq:15}
\end{eqnarray}
Thus, for a given experiment and set of parameters $s_i$,  we can 
construct a set of variables $X_i$ that are orthogonal linear 
combinations of the original variables; 
$X_1$  is the most accurately determined
parameter, $X_2$ is the next most accurately determined parameter 
and so on. We call the variables ${\bf X}$ the principal components
of the experiment, and refer to $X_7$, $X_6$ {\it etc} as `low-order'
components because they are poorly constrained
(smallest eigenvalues) and $X_1$, $X_2$,
{\it etc} as `high-order' components.

\medskip

\begin{figure*}
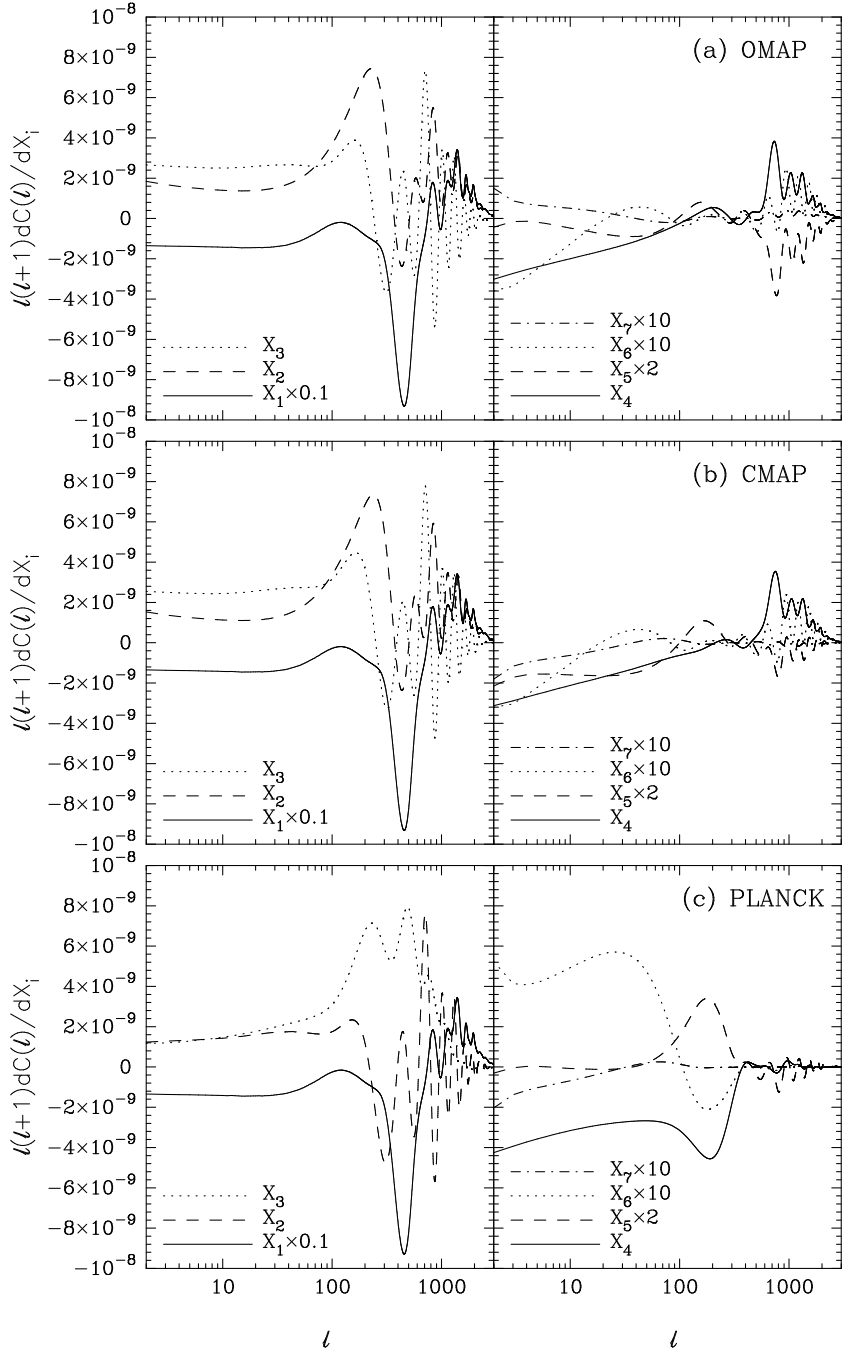


\vskip 7.1 truein

\includegraphics{pgfig6a.ps}

\includegraphics{pgfig6b.ps}

\includegraphics{pgfig6c.ps}

\caption{Derivatives of $C_\ell$ with respect to the principal
components listed in Table \ref{tab3} for the OMAP, CMAP and Planck experiments.
The principal components have been ordered so that $X_1$ is the
most accurately determined ({\it i.e.} has the largest eigenvalue),
$X_2$ is the next most accurately determined and so on. The derivatives
of $C_\ell$ with respect to $X_1$ and  $X_5$--$X_7$ have been multiplied
by the factors indicated in each figure.}
\label{figure6}
\end{figure*}

We have computed the eigenvalues and principal components
for each set of experimental parameters listed in Table \ref{tab1}.
The results are given in Table \ref{tab3},  which lists the components
of the projection vector $U^\dagger$ relating the principal
components to the original variables $s_i$. (Note that the
numbers in Table \ref{tab3} are computed assuming that the variables
$s_i$ have zero mean.) The derivatives of $C_\ell$ with respect
to the principal components for each experiment are plotted in
Figure \ref{figure6}. We make the following points concerning this analysis:

\noindent
[1] As in many aspects of multivariate analysis, certain results
depend on the scaling of the parameters. The principal components in
Table \ref{tab3} are not unique and depend on the relative scalings of the
variables $s_i$, as well as on the functional form of the variables
({\it e.g.} whether we use ${\rm log}\; \omega_b$ instead of
$\omega_b$, or $Q^2$ instead of $Q$). One must therefore be cautious
in assigning a physical interpretation to the principal components.
The point of view adopted here (in line with most applications of
principal component analysis, see {\it e.g.} Kendall 1975) is that the
principal components provide a computational tool for
assessing  whether a set of observational points ${\bf s_i}$
lie within a sub-space of the $n$-dimensional
hyperspace, $i = 1, \dots n$. ({\it i.e.} whether 
there are degeneracies among the physical parameters which are
unresolved by the experiment).

\noindent
[2] The standard deviation of the principal component $X_i$ is equal
to $\lambda_i^{-1/2}$. Thus, the eigenvalues have sometimes been used
to indicate the power of a CMB experiment ({\it e.g.} BET97). For
example, from Table \ref{tab3} we see that OMAP determines three
principal components to an accuracy of $\le 10^{-3}$, while CMAP and
Planck determine $4$ and $5$ components respectively to this accuracy
or better.  While this is true, what is important is how many {\it
physical} ({\it i.e.}  cosmologically interesting) parameters can be
determined to high accuracy rather than the number of linear
combinations like principal components that can be determined to high
accuracy.

\noindent
[3] The derivatives of $C_\ell$ with respect to the physical variables
plotted in Figure \ref{figure5} can be broadly grouped into two
categories: those that depend on inflationary variables, $n_s$, $n_t$
and $r$ that are largest at low multipoles, and those that depend on
physical densities, $\omega_b$, $\omega_c$, $\omega_\Lambda$, that
have significant amplitudes to $\ell \sim 2000$---$3000$. Cosmic
variance sets a fundamental limit to the accuracy with which the
variables $n_s$, $n_t$ and $r$ can determined from observations of the
CMB, whereas the cosmic densities can be determined to high precision
if $C_\ell$ is measured accurately at high multipoles. Thus, the power
of an experiment can be gauged by the extent to which the principal
components (particularly those with the largest variances) mix
inflationary variables with cosmic densities. Qualitatively, this can
be assessed from Figure \ref{figure6}. For OMAP and CMAP, the $C_\ell$
derivatives with respect to the low-order principal components are
oscillatory and have a significant amplitude at multipoles $\ell \sim
100$ --$2000$. This tells us immediately that the low-order principal
components involve the cosmic densities at some level and hence that
they will be partially degenerate with inflationary variables, which
is obviously unsatisfactory if we want to measure the cosmic densities
to high accuracy.

\noindent
[4] The principal components for a Planck-type experiment are nearly
equivalent to physical variables. In Table \ref{tab3}, we have placed an
asterisk next to each entry with an absolute value greater than $0.5$,
so that the reader can see easily which physical variables contribute
strongly to each principal component. For Planck, $X_1$, $X_2$, $X_3$,
$X_4$ and $X_5$ couple strongly to $\omega_b$, $\omega_c$, $Q$, $n_s$
and $\omega_\Lambda$ respectively. The two lowest order principal
components are made up predominantly of the two tensor parameters $r$
and $n_t$. Nevertheless, the degeneracies that remain dominate the
errors of the cosmological densities.  These remaining degeneracies
are, in a sense, fundamental, because they cannot be removed by making
more accurate observations of the linear CMB temperature anisotropies. 
The results for Planck shown in Table \ref{tab3} and Figure \ref{figure6} are very
close to those for an experiment limited by cosmic variance alone.
Accurate CMB {\it polarization} measurements on large scales could,
however, constrain the tensor component and so improve the accuracies
of the cosmic densities. Section \ref{sec:4.6} demonstrates how the accuracies
of the cosmic densities $\omega_b$ and $\omega_c$ are affected by
constraints on the tensor component.

In the rest of this Section, we investigate the degeneracies among
cosmological parameters as defined by the low-order principal
components. We conclude this Section by listing in Table \ref{tab4},
under the column headings `all pc', the $1\sigma$ errors in the
cosmological parameters derived from the Fisher matrix for the
spatially flat case.

For most of the parameters, the error estimates are
close to those listed in Table \ref{tab2}. The errors on $\omega_\Lambda$
in Table \ref{tab4} are obviously much smaller than those given in Table
\ref{tab2} because we are restricting to a spatially flat 
background model. The errors on $\omega_\Lambda$ will be discussed
in greater detail in Section \ref{sec:4.4}. The largest differences between
Tables \ref{tab2} and \ref{tab4} are for the errors in the tensor parameters
$r$ and $n_t$. The errors on these parameters are very large, 
so the first order expansion of $C_\ell$, on which the Fisher
matrix expression (4) is based, cannot necessarily 
be expected to give accurate
answers.

\vskip 0.2 truein

\begin{table}\label{tab4}
\bigskip
\centerline{\bf Table 4: $1\sigma$ errors in 
estimates of cosmological }
\centerline{\bf parameters (spatially flat universe)}
\medskip
\begin{center}
\begin{tabular}{ccccc} \hline \hline
\noalign{\medskip}
           &  \multicolumn{4}{c}{OMAP} \\
Parameter  &  all pc  & 1pc & 2pc & 3pc \\ \hline
$\delta \omega_b/\omega_b$        & $0.11$ & $0.056$ & $0.098$ & $0.10$ \\
$\delta \omega_c/\omega_c$        & $0.21$ & $0.080$ &  $0.19$ & $0.20$ \\
$\delta \omega_\Lambda$           & $0.15$ & $0.064$ & $0.14$  & $0.15$ \\
$\delta Q$                        & $0.014$& $0.0052$ & $0.0079$ & $0.011$ \\
$\delta r$                        & $0.81$  & $0.81$ & $0.81$ & $0.81$  \\
$\delta n_s$                      & $0.066$ & $0.030$ & $0.062$ & $0.063$  \\
$\delta n_t$                      & $0.745$ &   $0.73$ & $0.74$   & $0.74$ \\ \hline \hline
\noalign{\medskip}
           &  \multicolumn{4}{c}{CMAP} \\
Parameter  &  all pc & 1pc & 2pc & 3pc \\ \hline
$\delta \omega_b/\omega_b$  & $0.072$     & $0.024$  & $0.063$ & $0.064$ \\
$\delta \omega_c/\omega_c$  & $0.15\;\;$  & $0.044$ & $0.15$ & $0.15$ \\
$\delta \omega_\Lambda$     & $0.11$    & $0.034$ &  $0.11$ & $0.11$   \\
$\delta Q$                  & $0.0071$   & $0.0018$ &  $0.0038$ & $0.0041$ \\
$\delta r$              & $0.70$  & $0.70$ & $0.70$ & $0.70$   \\
$\delta n_s$            & $0.046$ & $0.0013$  & $0.044$ & $0.044$  \\
$\delta n_t$            & $0.71$ & $0.71$    & $0.71$ & $0.71$  \\ \hline \hline
\noalign{\medskip}
           &  \multicolumn{4}{c}{PLANCK} \\
Parameter  &  all pc & 1pc & 2pc & 3pc \\\hline
$\delta \omega_b/\omega_b$  & $0.0068$     & $0.0008$  & $0.0024$ & $0.0040$ \\
$\delta \omega_c/\omega_c$  & $0.0063$     & $0.0003$  & $0.0003$ & $0.0060$ \\
$\delta \omega_\Lambda$     & $0.0049$     & $0.0003$  & $0.0003$ & $0.0049$ \\
$\delta Q$                  & $0.0013$   & $0.0002$ &  $0.0006$ & $0.0006$ \\
$\delta r$                 & $0.49$  & $0.49$ & $0.49$ & $0.49$   \\
$\delta n_s$            & $0.0050$ & $0.0008$  & $0.0026$ & $0.0026$ \\
$\delta n_t$            & $0.57$ & $0.57$    & $0.57$ & $0.57$  \\ \hline \hline
\end{tabular}
\end{center}
\end{table}

\subsection{Degeneracies among cosmic parameters}\label{sec:4.3}

The columns in Table \ref{tab4} labelled `1pc', `2pc' and `3pc', list
the variances of the cosmological parameters that arise if we include
in the Fisher matrix analysis only the lowest order principal
component ($X_7$, column labelled `1pc'), the two and three lowest
order principal components ($X_6$ and $X_7$ in the column labelled
`2pc', and $X_5$, $X_6$ and $X_7$ in the column labelled `3pc'). {\it
For CMAP and OMAP, the two lowest order principal components account
for most of the variance of all the cosmological parameters listed in
the table}. Since we have specialized to a spatially flat universe,
the geometrical degeneracy described in Section \ref{sec:3} has been
explicitly removed from the analysis. Nevertheless, to a good
approximation, the parameters of Table \ref{tab4} lie on a two
dimensional surface within the 7-dimensional parameter space.  It is
the orientation of this surface with respect to the axes defined by
the parameters that accounts for the most of their variance.

\begin{figure*}
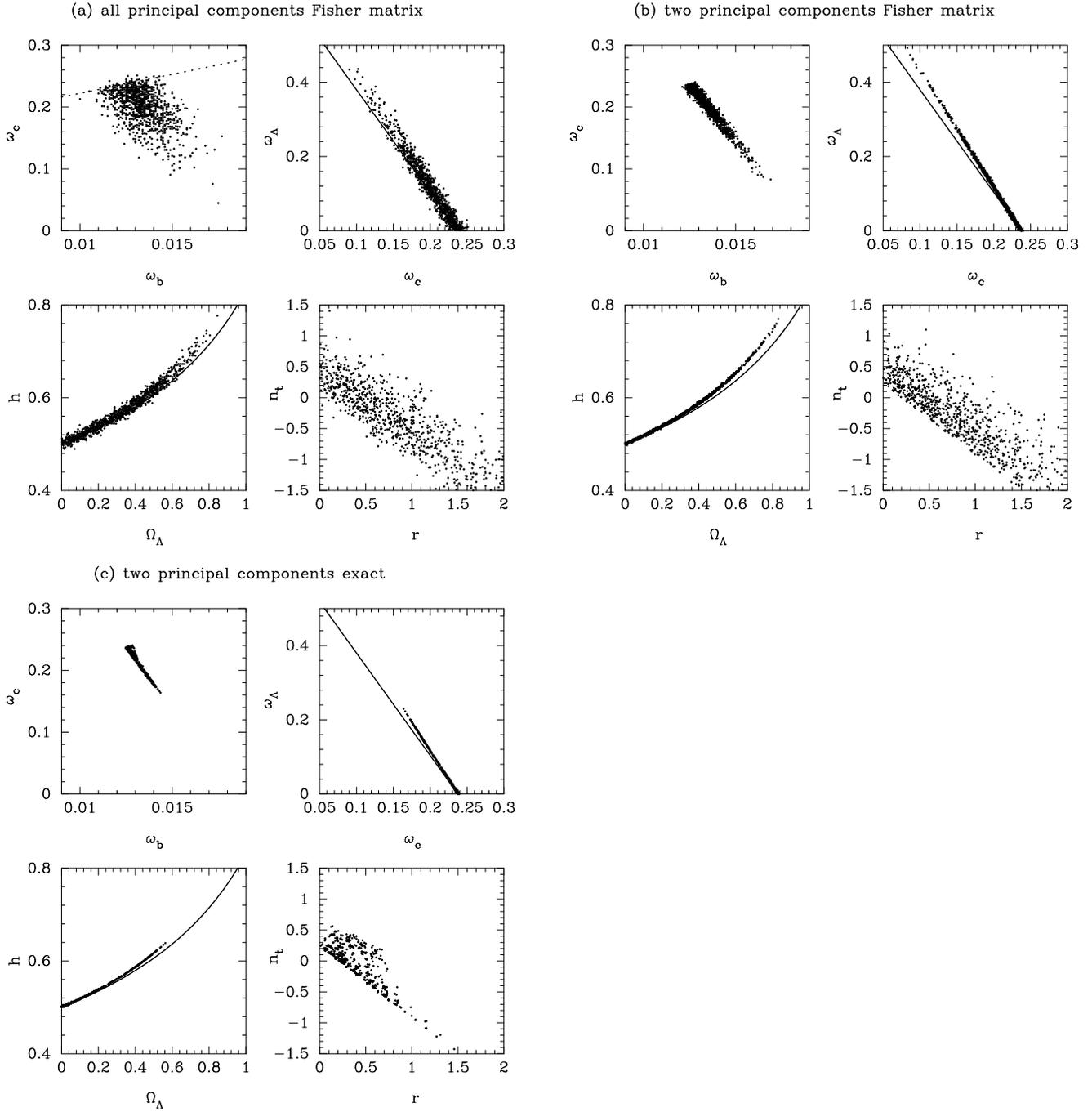


\vskip 7.5 truein

\includegraphics{pgfig7a.ps}

\includegraphics{pgfig7b.ps}
\includegraphics{pgfig7c.ps}
%{\hangindent=30pc \hangafter= 14
\caption{ Monte-Carlo simulations of parameter estimation as described
in the text showing correlations between various pairs of
parameters. We assume the experimental parameters for OMAP as given in
Table \ref{tab1}. In panel (a) we generate simulations using all
principal components, reproducing the variances and correlations
computed from the Fisher matrix. In panel (b) we generate simulations
using only the two lowest order principal components $X_6$ and $X_7$
(Table \ref{tab2}). In panel (c) we show the results of an exact
likelihood analysis applied to a two-dimensional grid of models in the
variables $X_6$ and $X_7$ (see Figure \ref{figure8} and the discussion
in the text). The solid lines in the $\omega_\Lambda$---$\omega_c$ and
$h$---$\Omega_\Lambda$ panels are computed from equations
(\ref{eq:23a}) and (\ref{eq:23b}). The dashed line in the
$\omega_c$--$\omega_b$ plane shows the degeneracy line given by a
constant height for the first Doppler peak as described in Section
\ref{sec:4.5}.}
%%\smallskip}
\label{figure7}
\end{figure*}

This is illustrated in Figure \ref{figure7}. In Figure \ref{figure7}a, we have generated
points in the 7-dimensional space of our cosmological parameters by
using all principal components as given in Table \ref{tab3} for the OMAP
experiment. The principal components are assumed to be Gaussian
distributed with dispersion equal to $1/\sqrt{\lambda_i}$.  We have made
one significant change to the Fisher matrix analysis of Section \ref{sec:2} in
that we have applied a constraint that the densities $\omega_i$ and
amplitude $r$ must be positive. This is evidently physically
reasonable for the $\omega_i$, required for $r$, 
and aids in the comparison with the exact calculations shown in
Figure \ref{figure7}c, to be described below.  We show correlations between
various pairs of parameters, but note in particular the very strong
correlations between $\omega_\Lambda$ and $\omega_c$ and between $h$
and $\Omega_\Lambda$. The points in these plots lie tightly on lines,
similar to the situation shown in Figure \ref{figure4}, which illustrated 
the geometrical degeneracy. As we will show below, the
strong correlations between $\omega_\Lambda$ and $\omega_c$ and $h$
and $\Omega_k$ have a similar physical explanation to the geometrical
degeneracy, though the degeneracy is not exact in this case.

Figure \ref{figure7}b is constructed in exactly the same way as Figure
\ref{figure7}a, except that we include only the two lowest order
principal components. As expected from the entries in Table
\ref{tab4}, by incorporating only the two lowest order principal
components, we reproduce the degeneracy lines in Figure \ref{figure7}a
which account for most of the variance of the physical parameters.
The higher order principal components contribute to the scatter around
these degeneracy lines but do not contribute much to the total
variance.

\begin{figure}

\vskip 5.5 truein

\includegraphics{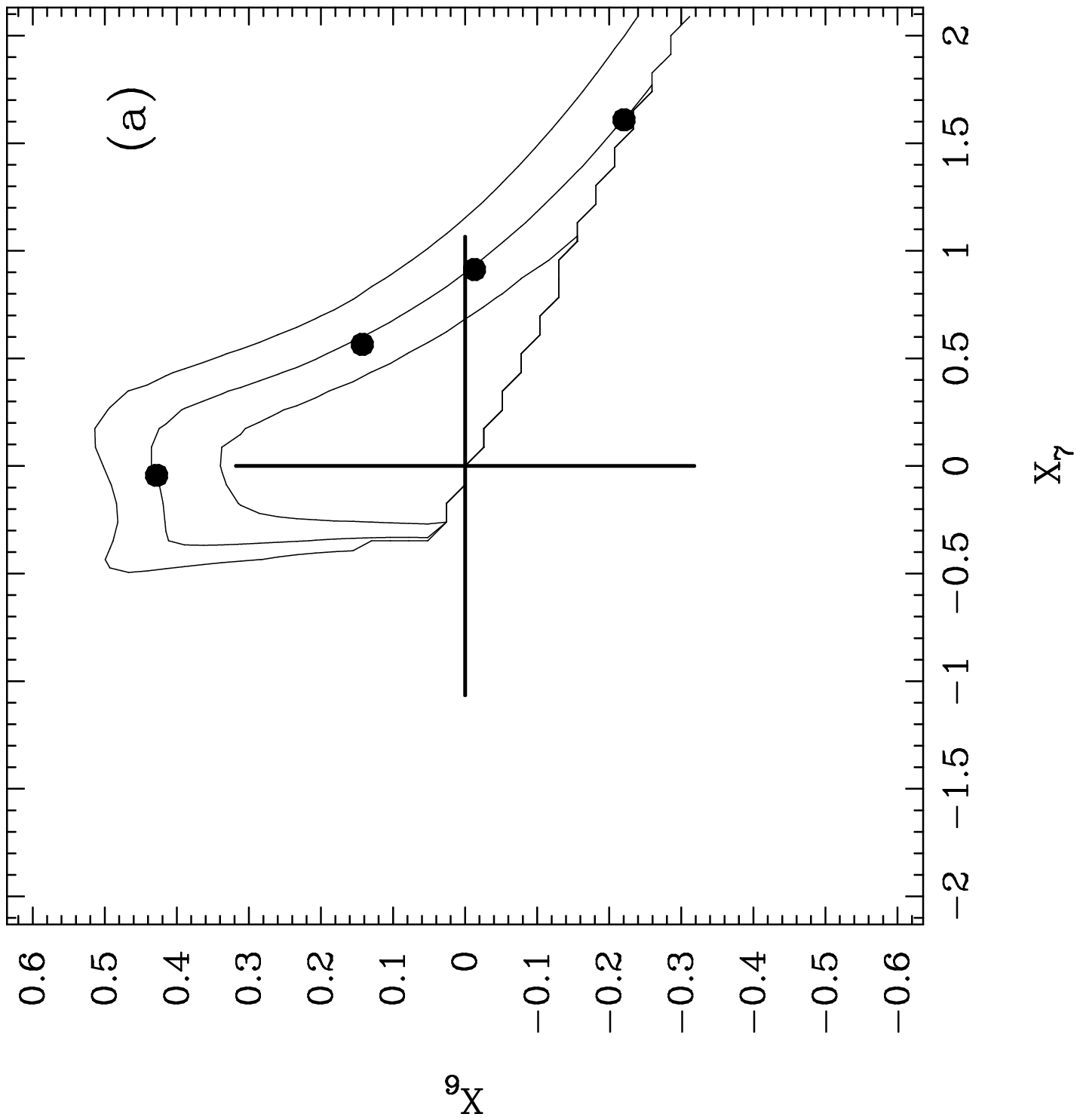}

\includegraphics{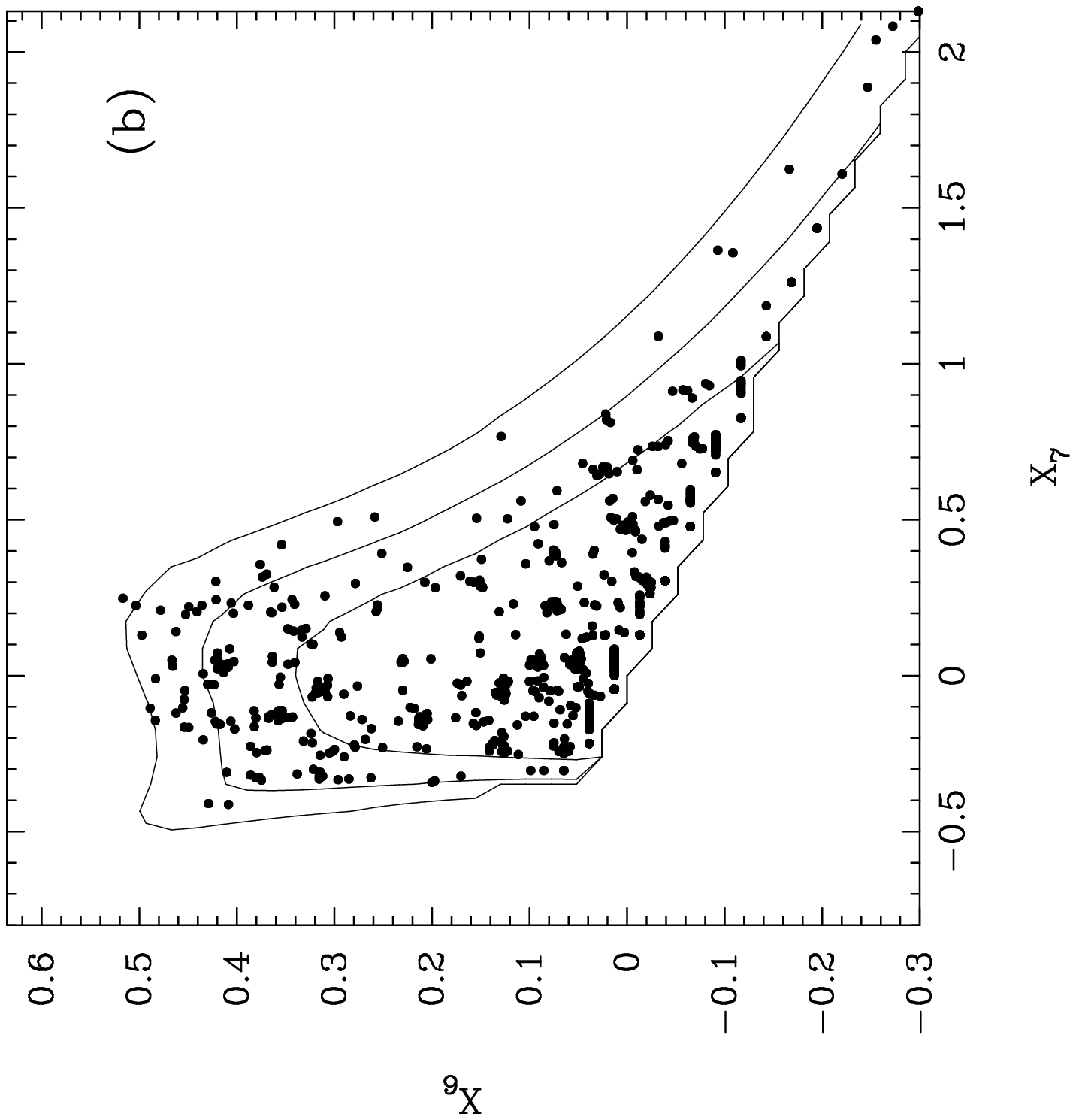}

\caption{Likelihood ratio contours in the $X_6$-$X_7$ plane for the
OMAP experiment ($1$, $2$ and $3\sigma$ contours computed as described
in the caption to Figure 3). The cross in Figure \ref{figure8}a shows
the $1\sigma$ errors derived from the Fisher matrix (computed from the
eigenvalues listed in Table \ref{tab3}).  The four filled circles show
the locations of the four models plotted in Figures \ref{figure10} and
\ref{figure11} below. Figure \ref{figure8}b shows the maximum
likelihood estimates of $X_6$ and $X_7$ for the Monte-Carlo models
described in the text. These values of $X_6$ and $X_7$ were used to
compute the values of the physical parameters plotted in Figure
\ref{figure7}c.}
\label{figure8}
\end{figure}

Table \ref{tab4} shows that the errors on some physical variables are
large ({\it e.g.} $\omega_\Lambda$, $r$, $n_t$).  As a consequence,
the variances of the low order principal components are also large and
so this raises a question concerning the validity of the Fisher matrix
approach to parameter errors, since this relies on small variations
$\Delta s_i$ of the parameters around those of the target model. To
test the validity of the Fisher matrix approach, we have computed an
exact version of Figure \ref{figure7}b using maximum likelihood.  We
define a $400\times 400$ grid in the principal components $X_6$ and
$X_7$ and compute scalar and tensor power spectra using the CMBFAST
code. The ranges of $X_6$ and $X_7$ are illustrated in Figure
\ref{figure8}a. The contours in Figure \ref{figure8}a show $1$, $2$
and $3\sigma$ contours for the likelihood ratio computed from equation
(\ref{eq:9}) with $\ell_{max} = 1000$. The jagged lower boundary of
the contours in the figures is imposed by positivity constraints on
the physical parameters, {\it i.e.} we cannot allow negative values of
$\omega_b$, $\omega_c$ and $r$ and (in this example) we do not allow
negative values of $\omega_\Lambda$.  The cross in
Figure~\ref{figure8}a shows the $\pm 1\sigma$ errors on $X_6$ and
$X_7$ computed from the eigenvalues listed in Table \ref{tab3}; note
that the scales of $X_6$ and $X_7$ in this figure have been chosen so
that the error contours computed from the Fisher matrix are circular.
Qualitatively, the likelihood errors are of about the same size as
those computed from the eigenvalues, but the distributions computed from 
the likelihood function are more sharply peaked than a bivariate
Gaussian distribution. This, together with the positivity constraints
which further narrows the distribution, means that the errors in $X_6$
and $X_7$ computed from the likelihood ratios are smaller than those
computed from the eigenvalues of the Fisher matrix.

Figure \ref{figure8}b shows a series of Monte-Carlo simulations. For
each point plotted in the figure we generated a simulated power
spectrum $C_\ell$ assuming Gaussian errors with variance given by
equation (\ref{eq:5a}) with the parameters of the OMAP experiment
listed in Table \ref{tab2}. For each simulated $C_\ell$ spectrum, we
computed the maximum likelihood values of $X_6$ and $X_7$ by comparing
with the grid of numerical models computed with CMBFAST. (In fact, we
reduce the effect of the finite grid using bi-linear interpolation to
compute the maximum likelihood values of $X_6$ and $X_7$.) Each point
in Figure \ref{figure8}b is plotted at the best fitting value of $X_6$
and $X_7$, and the distribution of points follows the likelihood
contours extremely well showing that our computations are self
consistent. From the values of $X_6$ and $X_7$, we compute the
physical parameters $s_i$ and so we can plot an analogue of Figure
\ref{figure7}b using parameters computed from an exact likelihood
approach. This is shown in Figure \ref{figure7}c.  The general trends
between the variables are similar to those shown in Figures
\ref{figure7}a and 7b, but the ranges are significantly narrower
showing that the Fisher matrix approach overestimates the errors of
the physical parameters. The Fisher matrix approach does give
qualitatively correct results (which is impressive given the large
errors in some of the physical variables) but evidently does not give
precise error estimates.

 Furthermore, since much of the variance in all of the physical
parameters comes from variations of the two lowest order principal
components, the Fisher matrix approach {\it will overestimate the
errors in apparently well determined parameters such as $\omega_b$},
since these are dominated by the variances of the low-order principal
components. This is also illustrated by the entries in Table
\ref{tab5} which is a tabular version of Figure \ref{figure7}. The
second and third columns list the variances of the physical parameters
(including $h$ and $\Omega_\Lambda$) computed from all principal
components and the two lowest order principal components; these
numbers differ from the corresponding numbers for OMAP in Table
\ref{tab4}, only because we have applied positivity constraints on $r$
and the densities $\omega_\Lambda$, $\omega_c$ and $\omega_b$.  The
fourth column shows the $1\sigma$ errors in the physical parameters
computed from the Monte-Carlo simulations plotted in Figure
\ref{figure7}c. Table \ref{tab5} shows that the Fisher matrix
overestimates the errors in many of these parameters by a factor of
two or so.

\begin{table}\label{tab5}
\bigskip
\centerline{\bf Table 5: $1\sigma$ errors in estimates of 
cosmological}
\centerline{\bf parameters: comparison
of Fisher matrix with} 
\centerline{\bf  exact likelihood including positivity constraints.}
\begin{center}
\begin{tabular}{|cc|c|c|} \hline \hline
           &  \multicolumn{3}{c|}{OMAP} \\ 
           & \multicolumn{2}{c|}{Fisher} & Likelihood \\ 
Parameter  &  all  & 2pc & 2pc  \\ \hline
$\delta \omega_b/\omega_b$        & $0.082$ & $0.061$ & $0.032$ \\
$\delta \omega_c/\omega_c$        & $0.13$  & $0.12$ &  $0.071$ \\
$\delta \omega_\Lambda$           & $0.094$ & $0.091$ & $0.052$ \\
$\delta Q$                        & $0.013$ & $0.0047$ & $0.0022$ \\
$\delta r$                        & $0.55$  & $0.55$ & $0.24$  \\
$\delta n_s$                      & $0.045$ & $0.040$ & $0.022$ \\
$\delta n_t$                      & $0.58$ &   $0.58$ & $0.32$  \\
$\delta h/h$                      & $0.11$ &   $0.11$ & $0.064$ \\
$\delta \Omega_\Lambda$           & $0.19$ &   $0.18$ & $0.15$   \\ \hline
\end{tabular}
\end{center}
\end{table}

The example shown in Figure \ref{figure7}c and Table \ref{tab5} shows how the principal
component analysis can be used to simplify a likelihood calculation of
the errors in cosmological parameters, so improving on the Fisher
matrix approximation. In general, the cosmological parameters defining
a model ${\bf s}_i$ are highly degenerate with respect to each other
and so the variances of the ${\bf s}_i$ are dominated by a small
number of principal components. In the examples given in Table \ref{tab4}, the
two lowest order principal components account for most of the variance
of the physical parameters. One can therefore use the principal
components of the Fisher matrix to reduce the dimensionality of the
parameter set to say two or three dimensions. It is then feasible to
evaluate the $C_\ell$'s numerically on a grid in this reduced
parameter space using a fast code such as CMBFAST. The numerically
computed power spectra can then be used to calculate the errors on the
physical parameters via a likelihood analysis in the reduced
dimensional space. This procedure, although time consuming, is much
faster than attempting to calculate the $C_\ell$'s numerically on a
grid in all of the physical parameters defining the model (7
dimensions in the simplified example given above, 10 or more
dimensions in the more realistic models described by BET97 and ZSS97).

Another consequence of the strong degeneracies among physical
parameters is that the variance estimates of some parameters
are extremely sensitive to imposed constraints. For example, if we
restrict the range of possible models to those that have negligible
tensor mode contribution  ({\it i.e.} we constrain
the parameter $r$ to be zero) then this has a large effect on the
lowest order principal components and hence on the variances of most
of the physical parameters. This point will be described in more
detail in Section \ref{sec:4.6}.

\subsection{Location of Doppler peaks and degeneracies for 
spatially flat models}\label{sec:4.4}

In Section \ref{sec:3.2} we analysed the geometrical degeneracy between
the parameters $\omega_k$ and $\omega_\Lambda$.
In this section, we analyse a similar degeneracy for spatially
flat models, {\it i.e.} we restrict to $\omega_k=0$ and derive the
relationship between $\omega_\Lambda$ and $\omega_c$ (or $\omega_m$)
that leads to a nearly identical location of the 
Doppler peaks as a function of multipole. However, since the 
Doppler peak structure (heights and relative amplitudes of the peaks)
depends on the matter content of the Universe, the resulting
degeneracy is not exact. As we will see below, the locations of
the Doppler peaks can explain the  strong correlations in the
$\omega_\Lambda$--$\omega_c$  and $h$--$\Omega_\Lambda$ plots
shown in Figure \ref{figure7}a. However, a  precise experiment probing
high multipoles such as Planck can break these degeneracies
to high accuracy ({\it cf} Table \ref{tab4}).

The comoving wavenumber of the m'th Doppler maximum or 
minimum is given approximately by 
\begin{eqnarray}
 k_m r_s(a_r) &= & m \pi   \label{eq:16}
\end{eqnarray}
(Hu and Sugiyama 1995), 
where $r_s$ is the sound horizon at recombination as defined
in Section \ref{sec:3.2}. The sound horizon is given by,
\begin{eqnarray}
 r_s &= & {c \over \sqrt 3} {1 \over H_0 \Omega_m^{1/2}}
\int_0^{a_r}  {da \over ( a + a_{equ})^{1/2}}
{1 \over (1 + R)^{1/2}} \nonumber \\
 &=& {c \over \sqrt 3} {a_r^{1/2} 
\over H_0 \Omega_m^{1/2}}
I_s( \omega_m, \omega_b)\, .\label{eq:17}
\end{eqnarray}
The expression on the right
hand side of this equation defines the integral $I_s(\omega_m, 
\omega_b)$ which is used in equation (\ref{eq:22}) below. The parameter
 $a_{equ}$ is the scale factor at the time that matter
and radiation (including massless neutrinos) have equal density,
and $R = (3 \rho_b/4 \rho_\gamma)$.   Thus,
\begin{eqnarray}
a_{equ}^{-1} & = & 24185\left ({1.6813 \over 1+\eta_\nu} \right )\omega_m,
\qquad R = 30496 \omega_b a\, . \label{eq:18}
\end{eqnarray}
where $\eta_\nu$ denotes the relative densities of massless
neutrinos and photons, $\rho_\nu = \eta_\nu \rho_\gamma$, and is
equal to $0.6813$ for three massless neutrino flavours. 
Evaluating the integral in equation (\ref{eq:17}) gives,
\begin{eqnarray}
r_s & =&  {4000 \over \omega_b^{1/2}} {a_{equ}^{1/2} \over (1 + \eta_\nu)^
{1/2}} \nonumber \\
 &{\rm ln}& \left [ { (1 + R(z_r))^{1/2} + (R(z_r) + R_{equ})^{1/2}
\over 1 + \sqrt{R_{equ}} } \right ] \;\; {\rm Mpc} \, .\label{eq:19}
\end{eqnarray}
We define the redshift of recombination $z_r$ as the redshift
at which the optical depth to Thomson scattering is unity,
assuming the standard recombination history (Peebles 1968). A
useful fitting formula for $z_r$ is given by Hu and Sugiyama
(1996),
\begin{eqnarray}
z_r & = &1048[ 1 + 0.00124 \omega_b^{-0.738} ] 
[ 1 + g_1 \omega_m^{g_2}]\, , \label{eq:20}\\
g_1 & = &  0.0783 \omega_b^{-0.238} 
[ 1 + 39.5 \omega_b^{0.763} ]^{-1}\, , \nonumber \\
g_2 &=&  0.560 [ 1 + 21.1 \omega_b^{1.81} ]^{-1} \, . \nonumber
\end{eqnarray}
The locations of the Doppler peaks are therefore given by
\beglet 
\begin{eqnarray}
&& \ell_m  \approx m \pi {d_A(z_r) \over r_s}, \label{eq:21a}
\end{eqnarray}
where $d_A$ is the angular diameter distance to the last scattering
surface
\begin{eqnarray}
&& d_A   = {c \over H_0 \Omega_k^{1/2}} \; {\rm sinh} ( \omega_k^{1/2}
y)\, ,
 \label{eq:21b} 
\end{eqnarray}
\endlet 
and $y$ is given by equation (\ref{eq:8b}). In fact, equation
(\ref{eq:21a}) is approximate because the exact numerical coefficient
relating $\ell_m$ and $d_A/r_s$ depends on the projection of the
three-dimensional temperature power spectrum to a two-dimensional
angular power spectrum. The exact relation depends therefore on the
shape of the primordial power spectrum and on the Doppler peak number
$m$ (see {\it e.g.} Hu and White 1996).  From the numerical
computations of $C_\ell$ we find that the location of the maximum of
the first Doppler peak for a scale-invariant primordial spectrum of
scalar fluctuations is well approximated by the expression
\begin{eqnarray}
 \ell_D & \approx &   0.746 \pi \sqrt 3 (1 + z_r)^{1/2}{ \omega_m^{1/2}
\over \omega_k^{1/2} } { {\rm sinh} [ \omega_k^{1/2} y ] \over 
I_s ( \omega_m, \omega_b)}.   \label{eq:22}
\end{eqnarray}

\begin{figure}

\vskip 3.4 truein

\includegraphics{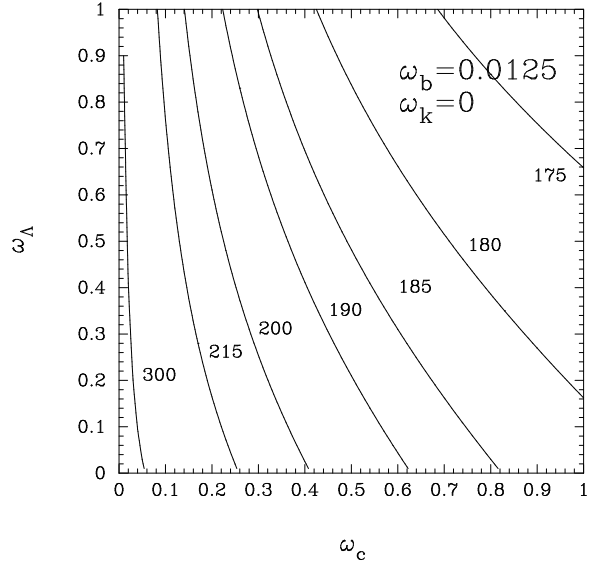}
\noindent
\caption{The position of the first Doppler peak, $\ell_D$, for a
spatially flat universe ($\omega_k = 0$) and assuming $\omega_b =
0.0125$.  The lines show contours of constant $\ell_D$, computed from
equation (\ref{eq:23a},\ref{eq:23b}), plotted as a function of
$\omega_\Lambda$ and $\omega_c$. The value of $\ell_D$ is given next
to each contour.}
\label{figure9}
\end{figure}

Equation (\ref{eq:22}) is valid for negatively curved
models, though we specialize to spatially flat models in
this section. To derive degeneracy lines for spatially flat models, we
take the limit $\omega_k \rightarrow 0$ in equation (\ref{eq:22}). For a fixed
value of $\omega_b$, the condition of fixed $\ell_D$ in equation (\ref{eq:22}) 
then leads to a relation between $\omega_\Lambda$ and $\omega_c$.
Figure \ref{figure9} shows the degeneracy lines for a spatially flat universe
with the baryon density of our target model ($\omega_b = 0.0125$).

As in Section \ref{sec:3.3}, for small variations of the parameters
around those of the target model, we can derive the degeneracy lines
by differentiating $\ell_D$. We do not expect the degeneracy lines
derived in this way to be exact because the degeneracies involve the
entire Doppler peak structure accessible to the experiment, not just
the position of the first Doppler peak. However, provided the baryon
density is determined to high precision from the experiment ({\it
i.e.} $\omega_b$ in equation \ref{eq:22} is held fixed), we would
expect that the positions of Doppler peaks would define the degeneracy
lines in the $\omega_c$--$\omega_\Lambda$ plane quite
accurately. These are given by 
\beglet
\begin{eqnarray}
&& \omega_c =   (\omega_c)_t + b \omega_\Lambda, \quad 
\qquad  b =  - { (\partial \ell_D/ \partial \omega_\Lambda)_t
 \over ( \partial \ell_D/\partial \omega_c)_t }\, , \label{eq:23a} 
\end{eqnarray}
where the subscript $t$ denotes that quantities evaluated assuming
the parameters of the target model. For our spatially flat target model, 
$b = -0.368$. As in the derivation of 
equation (\ref{eq:12}), the Hubble constant
is degenerate with $\Omega_\lambda$ according to 
\begin{eqnarray}
&&  h  =  { h_t  \over ( 1 - (1 + b) \Omega_\Lambda)^{1/2} }.
\label{eq:23b} 
\end{eqnarray}
\endlet 
The solid lines in the $\omega_\Lambda$--$\omega_c$ and
$h$--$\Omega_\Lambda$ panels in Figure \ref{figure7} are computed from
equations (\ref{eq:23a},\ref{eq:23b}). As expected, they give a good
approximation to the degeneracy lines, but the agreement is not exact.

\subsection{Height of the first Doppler peak and parameter degeneracies} \label{sec:4.5}

\begin{figure}

\vskip 3.0 truein

\includegraphics{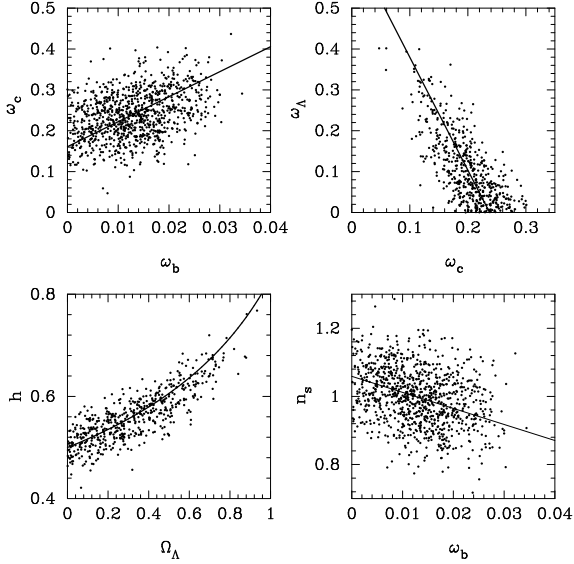}

\caption{Simulations of parameter estimation for an 
experiment sampling only the first Doppler
peak. We assume that the experiment is cosmic variance limited over
the multipole range $2 \le \ell \le 300$. The solid lines in the
$\omega_\Lambda$--$\omega_c$ and $h$--$\Omega_\Lambda$ planes
show lines of constant peak location as given by equations (\ref{eq:23a},)\ref{eq:23b}. The
dashed lines in the $\omega_c$--$\omega_b$ planes and in the $n_s$--$\omega_b$
plane show lines of constant Doppler peak height as given by equation \ref{eq:24a}. }
\label{figure11new}
\end{figure}

In analogy with the location of the Doppler peaks, the height of the 
first Doppler peak can provide some further insight into parameter 
degeneracies. The scalar component of the power spectrum can be
roughly approximated by the following expression
\beglet 
\begin{eqnarray}
&& {\ell(\ell+1) C_\ell \over 6 C_2} \approx {\ell (\ell +1 ) \over 6} 
{\Gamma (\ell + {(n_s - 1) \over 2})
\over \Gamma (\ell + {(5- n_s) \over 2})} {\Gamma ({(9 - n_s) \over 2})
\over \Gamma ({(3 + n_s) \over 2})} \nonumber \\
&&
+ {A(\omega_b, \omega_c, n_s) \over [1 + ({\ell_D \over 8 \ell})^{3/2}]}
{\rm exp}\left 
[ -{(\ell - \ell_D)^2 \over 2 (\Delta \ell_D)^2} \right ]\, , \label{eq:24a} 
\end{eqnarray}
where
\begin{eqnarray}
&&  \quad \Delta\ell_D = 0.42 \ell_D, \\
&& {\rm ln} A(\omega_b, \omega_c, n_s) \approx 4.5(n_s -1 ) + a_1 + a_2 \omega_c^2
+ a_3 \omega_c +  \\
&& \qquad \qquad a_4 \omega_b^2 +  a_5 \omega_b + a_6 \omega_b
\omega_c \, ,  \label{eq:24b}
\end{eqnarray}
\begin{eqnarray}
&& a_1 = 2.376, a_2 = 3.681, a_3 = -5.408, \nonumber \\
&& a_4 = -54.262, a_5 = 18.909, a_6 = 15.384. \nonumber 
\end{eqnarray}
\endlet 
and $\ell_D$ is given by equation (\ref{eq:22}) The first term
in equation (\ref{eq:24a}) is the usual Sachs-Wolfe expression for a
power-law fluctuation spectrum (see {\it e.g.} Bond 1996). The second
term is a fit to the first Doppler peak determined from a grid of
CMBFAST computations of $C_\ell$.

The main parameter degeneracies for an experiment sampling only the 
multipole range $\ell \simlt 300$-$400$ will be determined by the
properties of the first Doppler peak. Thus, the Doppler peak position
$\ell_D$ will determine the degeneracy between $\omega_\Lambda$, $\omega_c$ 
and $\omega_b$ (as described in the previous section) and the Doppler
peak height will determine the degeneracies among $\omega_b$, $\omega_c$
and $n_s$. This is illustrated in Figure \ref{figure11new}, which shows a similar plot
to Figure \ref{figure7}a, but for an experiment that is cosmic variance limited
for the multipole range $2 \le \ell \le 300$. The conditions of constant
peak height and location provide a good description of the parameter
degeneracies in this case.

The situation is more complicated for a MAP or Planck type experiment
because they are sensitive to the subsidiary Doppler peak structure.
This is illustrated by the $\omega_c$--$\omega_b$ scatter plot in
Figure \ref{figure7}a. To a reasonable approximation, the 
subsidiary Doppler peaks
pin down the value of $\omega_b$ to high accuracy and so the first
principal component is almost vertical in the $\omega_c$--$\omega_b$
plane (in fact it is tilted to a slightly negative slope to reduce
the scatter in the Doppler peak locations). In this case, the condition
of constant first Doppler peak height (plotted as the dashed line)
determines the scatter in the orthogonal direction
to that defined by the first principal component.

\subsection{Constraints on the Tensor Component}\label{sec:4.6}

\begin{figure*}
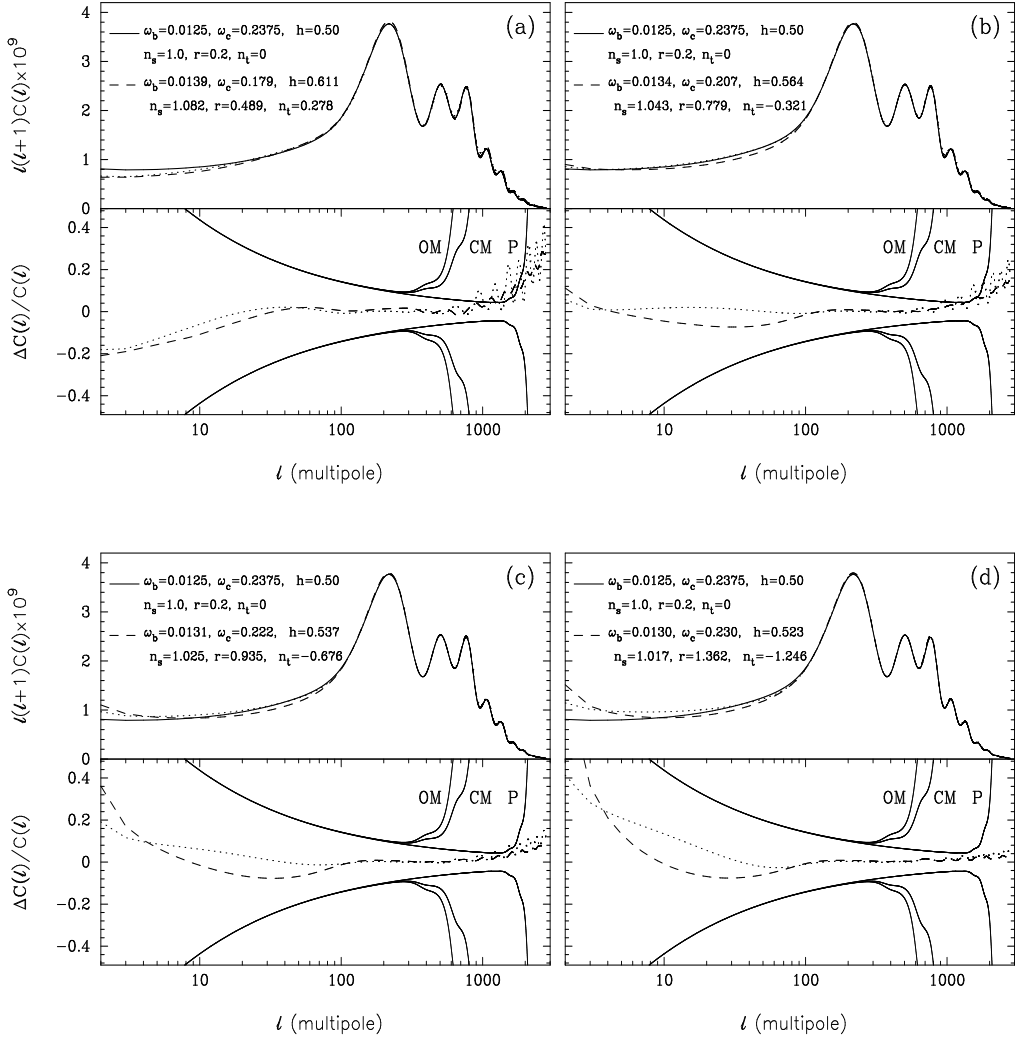


\vskip 2.5 truein

\includegraphics{pgfig10a.ps}

\includegraphics{pgfig10b.ps}

\vskip 3.3 truein

\includegraphics{pgfig10c.ps}

\includegraphics{pgfig10d.ps}

\caption{Temperature power spectra for the four models
plotted as the filled circles in Figure \ref{figure8}a. For these
models the two least well determined principle components $X_6$ and
$X_7$ lie on the $2\sigma$ contour in Figure \ref{figure8}a. The
physical parameters of these models are given in each panel. The solid
line in the upper panel in each figure shows the power spectrum of the
target model; the dashed line shows the exact computation of $C_\ell$
using CMBFAST for the parameters defined by $X_6$ and $X_7$.  The
dashed line shows $C_\ell$ inferred from the derivatives of $C_\ell$
(equation \ref{eq:15}). The lower panels show the residuals of these
curves with respect to the target model; dashed lines show the
residuals for the numerical computations of $C_\ell$ and dotted lines
show the residuals of the derivative approximation.  The lines
labelled $OM$, $CM$ and $P$ show the $1\sigma$ errors computed from
equation (\ref{eq:5a}) for the parameters of OMAP, CMAP and Planck
given in Table \ref{tab1}.}
\label{figure10}
\end{figure*}

Figure \ref{figure10} shows the temperature power spectra for the four
models indicated by the dots in Figure \ref{figure8}a. These models
lie on the $2\sigma$ contour in the plane defined by the lowest order
principal components. An experiment with the OMAP parameters can
therefore only marginally distinguish these models from our target
model. Figure \ref{figure10} is designed to illustrate the following
points:

\noindent
[1]  The parameters for these models are given in the upper panels. 
Notice that the the tensor parameters $n_t$ and $r$ vary by large 
amounts and that the most obvious differences with the power 
spectrum of the target model are at low multipoles. This is 
expected because the poorly determined tensor parameters carry high
weight in the lowest two principal components ({\it c.f.} Table \ref{tab4}).

\noindent
[2] The dotted lines in the upper and lower panels show the power
spectra computed from the first derivatives according to
equation~(\ref{eq:13}).  These differ significantly from the numerical
computations (shown as the dashed lines) because the variations in
$n_t$ and $r$, and of some other parameters, are so large that the
first order Taylor expansion breaks down. This is why the exact
likelihood analysis shown in Figure \ref{figure7}c differs from the
Fisher matrix approximation used to derive Figure \ref{figure7}b. The
exact analysis leads to tighter constraints on the physical parameters
because the first order approximation tends to underestimate the
differences in the power spectra.

\noindent
[3] Throughout this paper, we have allowed the parameters $n_s$, $n_t$ 
and $r$ to vary independently of each other. In other words, we have
not imposed any additional constraints on the physical mechanism which 
gave rise to the fluctuations, other than that they are Gaussian
characterised by power-law spectra. However,  wide classes
of inflationary models lead to  constraints between the scalar
and tensor component (see the comprehensive review by Lidsey {\it etal}
1997 and references therein). Such constraints can severely limit
the allowed  area in the plane defined by the lowest two principal
components,  and since these account for such a large proportion 
of the variance of {\it all} the physical parameters, 
the parameter errors  will be extremely
sensitive to these constraints.

We will investigate this last point in the remainder of this Section,
using as examples two limiting constraints. It is
useful first to review some key results concerning the scalar and
tensor perturbations predicted by inflationary models to motivate
these approximations.

Tensor perturbations, arising from quantum noise in the gravitational
wave degrees of freedom, obey a perturbation equation like that of a
free massless field, whereas the equation for scalar curvature
fluctuations have extra terms determined by the shape of the scalar
field potential $V(\phi)$, where $\phi$ is the inflaton field that
drives inflation.  The relationship between the two is therefore model
dependent, and can be quite complex if $V(\phi)$ is, {\it e.g.}, if
inflation involves more than one dynamically important scalar degree
of freedom. The essential features of inflation for one degree of
freedom can be understood using a Hamilton-Jacobi formulation (Salopek
and Bond 1990) in which the Hubble parameter $H(\phi )$ is treated as
a function related to $V(\phi)$ by
\begin{eqnarray}
   H^2 & =  & {8 \pi \over 3 m^2_p} \left [ {1 \over 2}
 \left ( {m^2_p \over 4 \pi}{\partial H \over \partial \phi} \right )^2 
+ V(\phi) \right ], \label{eq:25}
\end{eqnarray}
where $m_p$ is the Planck mass.  The amplitudes of the post-inflation
scalar and tensor fluctuation spectra, $P_S$ and $P_T$, can be
expressed in terms of the value of $H$ when a comoving wavenumber $k$
crosses the Hubble radius $k=Ha$ at time $\tau_k$: 
\beglet 
\begin{eqnarray}
&& P_S (k)  =  {1 \over  1 + q} {H^2(\tau_k) \over \pi m_p^2} e^{2u_s},
\label{eq:27a} \\ 
&&   P_T (k) =  { 8 H^2 (\tau_k) \over \pi m_p^2} e^{2u_t}\, . \label{eq:27b} 
\end{eqnarray}
\endlet 
The deceleration parameter $q$ is related to the slope of $H$ by
\begin{eqnarray}
   (1 + q)  & = &   {m^2_p \over 4 \pi}  \left ( {\partial {\rm ln} H 
\over \partial \phi} \right )^2 \,  \label{eq:26}
\end{eqnarray}
(e.g., Bond 1996). The parameters $u_s$ and $u_t$ (e.g., Bond 1994)
turn out to be small for power-law like potentials (but testing
whether they are non-zero is an important goal of CMB measurements).
Thus the ratio of tensor to scalar power is largely determined by
$(1+q)$, which measures the departure from pure exponential expansion
during inflation.

Taking logarithmic derivatives of the spectra $P_S$ and $P_T$ with
respect to wavenumber $k$ gives the scalar and tensor spectral indices
$n_s$ and $n_t$ introduced in Section \ref{sec:3},
\beglet
\begin{eqnarray}
&& n_s = 1 + 2 \left ( 1 +
{1 \over q} \right ) - {1 \over q} {m_p^2 \over 2 \pi} {\partial^2
{\rm ln} H \over
\partial \phi^2} + \epsilon_s , \label{eq:28a} \\
&&    n_t  =  2 \left ( 1 + {1 \over q} \right) + \epsilon_t, \label{eq:28b}
\end{eqnarray}
\endlet 
where the terms $\epsilon_s$ and $\epsilon_t$ are small corrections
arising from derivatives of $u_s$ and $u_t$. Thus the difference
$n_s-1-n_t$ is determined by the change in $q$ over the $k$ range
appropriate for large scale structure and the CMB, which turns out to
correspond to quite a narrow section of the inflaton potential. The
familiar `slow rollover' approximation assumes $q \approx -1$ in
equation (\ref{eq:25}), giving $H^2 \approx 8 \pi V/(3 m^2_p)$, but
not in equation~(\ref{eq:27a}) nor in
equations~(\ref{eq:28a},\ref{eq:28b}). Indeed for inflation to end,
the universe must enter a deceleration phase, so realistic potentials
must give departures from pure scale-invariant spectra, possibly large
but more plausibly small.

The ratio of the scalar and tensor components can also be computed in
terms of the inflation parameters defined above; e.g., if $n_s$ and $n_t$ are
assumed to be constant, 
\begin{eqnarray}
r  & = &   {C_2^T \over C_2^S} \approx 13.7(1 + q)e^{2(u_t-u_s)}e^{-0.15n_t}
e^{-1.1(n_s - 1 -n_t)} \label{eq:29} \\
&\approx &   {-6.8 n_t \over (1 - n_t)} e^{-0.15n_t} 
e^{-1.1(n_s - 1 -n_t)}  \label{eq:30}
\end{eqnarray}
(Bond 1996).  $r$ also explicitly depends upon parameters such as
$\omega_\Lambda$ and $\omega_k$.

There are two special cases we use to illustrate constraints on the
tensor component. The first is nearly uniform acceleration, (so
$\partial^2 \ln H /\partial \phi^2 \approx 0$), 
\beglet
\begin{eqnarray}
&&  n_t    \approx   n_s - 1, \  r \approx -6.8 {n_t \over (1 - n_t/2)} 
\approx 6.8 {(1 - n_s) \over (3-n_s)/2}.  \label{eq:31a} 
\end{eqnarray} 
Pure power law expansions have $q$ precisely constant whereas chaotic
inflation models with smoothly varying power law potentials have $q$ 
nearly uniform, but the deviations from $n_s-1\approx n_t$
are important to include if $n_t$ is small.  Equation
(\ref{eq:31a}), often approximated by $r \approx 7(1-n_s)$, has been
discussed by many authors (see {\it e.g.} Lidsey \etal 1997, and the
footnote in Section \ref{sec:3.1}). The second constraint we consider has
\begin{eqnarray}
&&  n_t    \approx  0,   \qquad   r \approx  0,  \label{eq:31b} 
\end{eqnarray} 
\endlet
 which applies to models such as `natural inflation' (Adams
\etal 1993) in which the inflaton begins near the maximum of an
extremely flat potential, so $1+q$ and the tensor mode are
exponentially suppressed. Since $d(1+q)/d\phi$ cannot then be
neglected, $n_s$ can be less than unity, possibly significantly
so. Similar behaviour can arise in inflationary models based on
supergravity ({\it e.g.}  Ross and Sarkar 1996). For specific classes
of models, one can refine the estimates of equations
(\ref{eq:31a},\ref{eq:31b}), for example by relating the corrections
$u_t-u_s$ to the spectral indices.

In Table \ref{tab6} we show how the errors in the cosmological
parameters change if we apply the constraints of equation
(\ref{eq:31a},\ref{eq:31b}). The table lists the $1\sigma$ errors
computed from the Fisher matrix including positivity constraints. The
columns labelled `no const.'  give the results when no constraints are
applied to the tensor spectrum, {\it i.e.} all parameters are allowed
to vary independently.  The entries under this heading for OMAP are
thus identical to the entries in the second column of Table
\ref{tab5}.  The columns labelled `$r = -7n_t$' show the results of
applying the constraint (\ref{eq:31a}) and those labelled `$r=0$' show
the result of applying the constraint (\ref{eq:31b}).

\begin{table*}\label{tab6}
\bigskip
\centerline{\bf Table 6: $1\sigma$ errors in estimates of 
cosmological parameters:}   
\centerline{ \bf effect of constraining the tensor component.}
\medskip
\begin{center}
\begin{tabular}{|c|ccc|ccc|ccc|} \hline\hline
           & \multicolumn{3}{c|}{OMAP} & \multicolumn{3}{c|}{CMAP} &
\multicolumn{3}{c|}{PLANCK} \\ 
Param.  & no const.  & $r=-7n_t$ & $r=0$& no const.
  & $r=-7n_t$ & $r=0$& no const.
& $r=-7n_t$ & $r=0$   \\ \hline
$\delta \omega_b/\omega_b$  & $0.082$ & $0.036$ & $0.046$ & $0.052$  
& $0.028$ & $0.030$ &  $0.0064$ & $0.0056$ & $0.0056$ \\
$\delta \omega_c/\omega_c$  & $0.13$  & $0.046$ & $0.044$ & $0.097$  
& $0.028$ & $0.031$& $0.0042$ & $0.0042$   & $0.0039$  \\
$\delta \omega_\Lambda$     & $0.094$ & $0.025$ & $0.026$ & $0.069$  
& $0.014$ & $0.020$& $0.0030$ & $0.0030$ & $0.0027$\\
$\delta Q$                  & $0.013$ & $0.007$ & $0.009$ & $0.0066$ 
& $0.0047$& $0.0050$&$0.0013$ & $0.0010$ & $0.0011$\\
$\delta r$                  & $0.55$  & $0.054$ &  ---    & $0.49$   
& $0.043$ &  ---     &  $0.33$ & $0.023$ &  --- \\
$\delta n_s$                & $0.045$ & $0.0077$ & $0.015$& $0.030$  
& $0.0061$& $0.0098$& $0.0049$ & $0.0032$ & $0.0042$ \\
$\delta n_t$                & $0.58$ &  $0.0077$ &  ---   & $0.56$   
& $0.0061$&  ---     &  $0.40$ & $0.0032$ & --- \\
$\delta h/h$                & $0.11$ &  $0.032$ & $0.033$ & $0.082$  
& $0.020$ & $0.028$ & $0.0045$ & $0.0045$ & $0.0041$ \\
$\delta \Omega_\Lambda$     & $0.19$ &  $0.079$ & $0.082$ & $0.16$   
& $0.049$ & $0.068$ &  $0.012$ & $0.012$ & $0.011$\\ \hline
\end{tabular}
\end{center}
\end{table*}

\vskip 0.1 truein

As expected (see point 3 above), imposing the constraints on a
OMAP-type experiment results in a significant reduction of the errors
on most parameters. Even for a CMAP-type experiment, the constraints
lead to large changes in the errors of some parameters, {\it e.g.} the
fractional error in $\omega_D$ decreases by about a factor of $3$, the
fractional error in $h$ decreases by a factor of $4$ and the error in
$\Omega_\Lambda$ decreases by a factor of between $2$ and $3$. For a
Planck-type experiment, however, imposing the constraints leads to
hardly any change in the parameter errors. As Table \ref{tab3} and
Figure \ref{figure6} show, for a Planck-type experiment, there is
almost a decoupling between the highly uncertain tensor parameters and
the high order principal components.

 The results of Table \ref{tab6} show that the errors on cosmological
parameters can be reduced by making certain specific assumptions
concerning the inflationary model. For OMAP and CMAP-type experiments,
the reduction in the errors is very significant, but at the expense of
overly restricting the inflation model. Allowing full freedom in how
$V(\phi)$ and thus $H(\phi)$ is structured can result in strong
variations in $q$, with the consequence that the power ratio
\ref{eq:27b} to \ref{eq:27a} and the indices $n_s(k)$ and $n_t(k) $
could be complex functions of the wavenumber. As might be expected, if
there is complete functional freedom, the errors on the cosmological
parameters become much larger (e.g., Souradeep \etal 1998). Such
inflation models with many features over the small regime of $V(\phi
)$ that the CMB can probe seem rather baroque, and constraining the
potential by invoking smoothness ``prior probabilities'' seems
perfectly reasonable. For example, a simple way that this has been
included in the past is to allow the spectral indices to ``run'' with
a logarithmic correction: $n_s(k) \approx n_s (k_s) + \alpha
\ln(k/k_s)$ (Kosowsky and Turner 1995, Lidsey \etal 1997).  Of course,
in presenting the results of CMB parameter estimation, the constraints
(if any) applied to the tensor mode must be spelled out precisely.

This analysis illustrates also that constraints on the tensor
component derived from CMB polarization measurements, could lead to a
significant improvement in the estimates of other parameters,
including the baryon and CDM densities.

\section{Strength, Significance and Correlated Errors in $C_\ell$}\label{sec:5}

The next generation of CMB satellites should lead to estimates of
$C_\ell$ with errors dominated by cosmic variance, 
\begin{eqnarray} 
&& {\Delta C_\ell
\over C_\ell} \approx \sqrt{ 2 \over (2\ell+1) f_{sky}}  \nonumber 
\end{eqnarray} 
for multipoles $\ell \sim 1000$ -- $2500$.  At $\ell \sim 1000$,
cosmic variance limits the $1\sigma$ accuracy of a measurement of
$C_\ell$ to about $3$ percent. At first sight, it might seem as though
there is little point in measuring $C_\ell$ (and subtracting
contaminating foregrounds) to a much higher accuracy than the limit
imposed by cosmic variance. However, such thinking would be
incorrect. The reason is related to the concepts of strength and
significance in statistics (see Press \etal 1996) and we illustrate
the point with the following simple example. Imagine that we have $N$
independent random variates, each drawn from a Gaussian with a mean
and dispersion of unity. From these $N$ random variates we can
estimate the mean
\begin{eqnarray} 
&& \mu = {1 \over N} \sum x_i, \nonumber 
\end{eqnarray} 
with variance
\begin{eqnarray} 
&& {\rm Var} (\mu) = {1 \over N}.  \nonumber 
\end{eqnarray} 
Thus, even though the variance on each individual point is unity, we
can determine the mean to arbitrarily high precision if we have a
large enough number of points, (${\rm Var}(\mu) \rightarrow 0$, in the
limit $N \rightarrow \infty$).  We can therefore differentiate between
two samples with slightly different means ($\delta \mu \ll 1$),
provided we have enough points in each sample.  In this example, each
point has very little {\it strength} (since the difference in the
means is much smaller that the variance) but in the limit of large
samples we can detect a small difference in the means to high
{\it significance}.

The estimation of cosmological parameters from measurements of the
power spectrum is analogous to the above example. The CMB satellites
will provide, via $C_\ell$, measurements of $1000$--$3000$ nearly
statistically independent numbers from which we wish to determine a
much smaller number (perhaps $10$--$20$) cosmological parameters.
Since we have a large number of independent points, it is possible to
measure parameter differences to high significance even though the
strength of each individual point is small. Furthermore, since the
number of cosmological parameters that we wish to measure is much
smaller than the number of independent multipoles, we must measure
each $C_\ell$ to an accuracy that is much smaller than the cosmic
variance; correlated systematic errors between the measured $C_\ell$
coefficients which introduces a difference in $\chi^2$ of the order
the number of {\it parameters} can lead to a significant bias in the
cosmological parameter estimates, the exact bias depending on the
nature of the correlations. We will expand on these points in the rest
of this Section.

\begin{figure*}
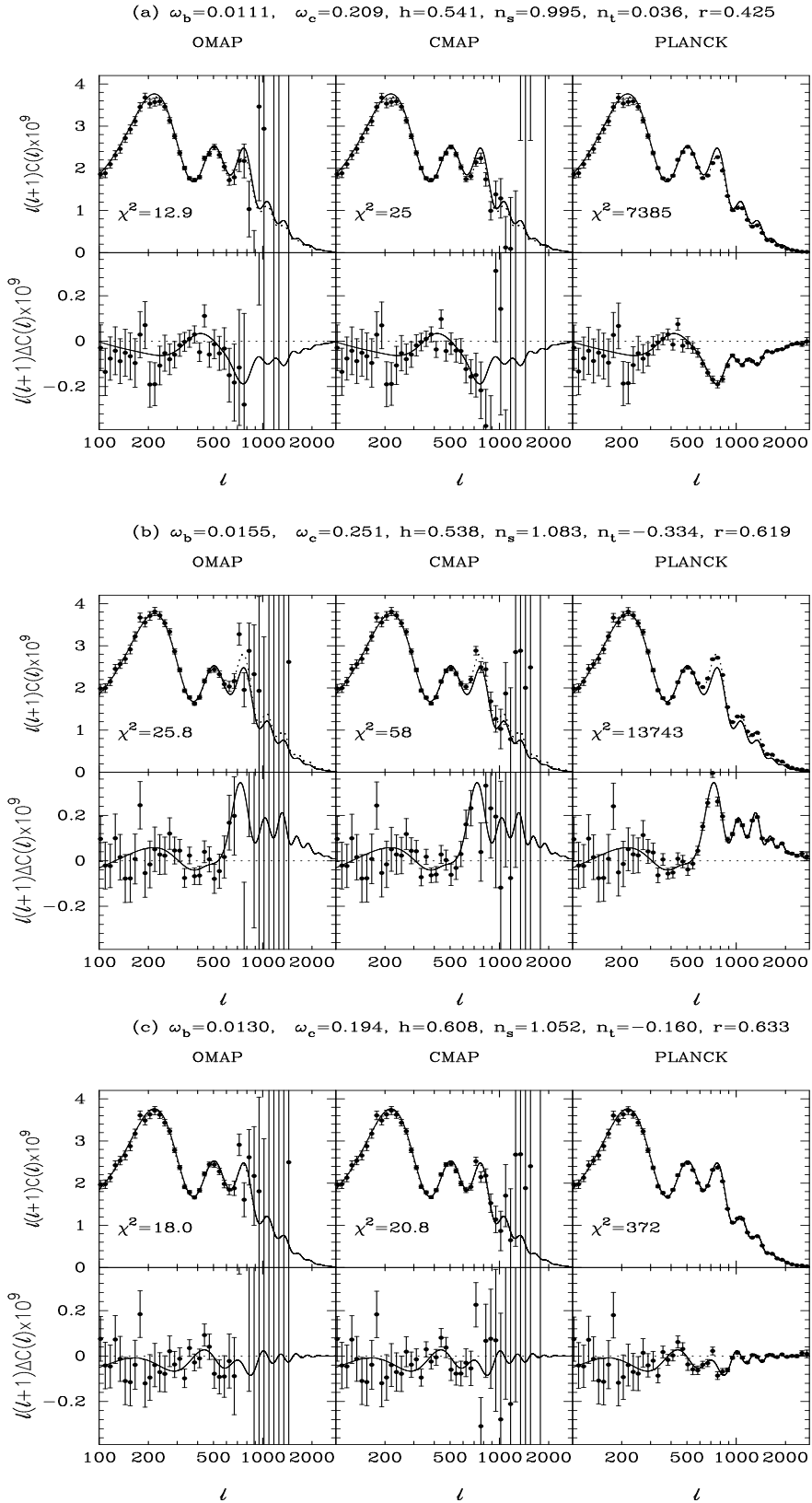


\vskip 8.9 truein

\vskip -0.2 truein
\includegraphics{pgfig11a.ps}

\vskip -0.1 truein
\includegraphics{pgfig11b.ps}

\vskip -0.1 truein
\includegraphics{pgfig11c.ps}

\vskip -0.1 truein

\caption{Temperature power spectra for three spatially flat models
that are distinguishable by OMAP from the target model at about the
$2\sigma$ level. The parameters for each model are listed at the top
of each figure and the temperture power spectrum computed for these
parameters is shown by the dashed lines in the upper panels. The power
spectrum of the target model is shown by the solid lines.  The filled
circles and error bars show a random realization of $C_\ell$ for each
model and $1\sigma$ errors assuming the experimental characteristics
of OMAP, CMAP and Planck as listed in Table \ref{tab1}. The points in
the lower panels show the residuals with respect to the target
model. The number in each panel gives $\chi^2$ per degree of freedom
for each experiment.  The points have been averaged in $100$ equally
spaced logarithmic intervals in $\ell$ over the range $3 \le \ell \le
3000$.}
\label{figure11}
\end{figure*}

Figure \ref{figure11} is designed to illustrated the concepts of
strength and significance as applied to measurements of $C_\ell$. The
dotted curves in the upper panels of the figure show $C_\ell$ for a
spatially flat CDM model with the parameters listed in the Figure
heading. The solid lines show the power spectrum of the spatially flat
target model defined in the previous Section. The parameters for the
three examples in this figure were chosen so that the $C_\ell$ curves
would be distinguishable by OMAP at about the $2 \sigma$ level; in
other words, we have chosen three pairs of spatially flat CDM models
that would be difficult to distinguish with OMAP alone even if there
were no significant systematic errors in the measurements. The three
examples have been chosen so that: in Figure \ref{figure11}(a) the
baryon and CDM densities each differ by about $12\%$ from those of the
target model, while the inflationary parameters $n_s$, $n_t$ and $r$
are almost the same as those of the target model; in Figure
\ref{figure11}(b), the baryon density differs by $24 \%$ from that of
the target model while $\omega_c$ differs by $6 \%$; in Figure
\ref{figure11}(c) the CDM density differs by $18 \%$ from that of the
target model while $\omega_b$ differs by $4\%$. In the models of
Figures \ref{figure11}(b) and \ref{figure11}(c), the large changes in
the densities have been balanced by relatively large changes in the
inflationary parameters \footnote{Note that to test inflationary
models at interesting levels, we would certainly want to determine the
spectral index $n_s$ to better than $\pm 0.05$ (see {\it e.g.} Liddle
1997 and references therein)}.  The points in each figure show a
simulated $C_\ell$ spectrum for each experiment, assuming the dotted
curve for $C_\ell$, and with errors computed from equation
(\ref{eq:5a}). The lower panels show the residuals relative to the
target model.

The number labelled $\chi^2$ in each panel gives the likelihood ratio
$\chi^2 = - 2 ln( {\cal L}/{\cal L}_{max})$, computed from
equation~(\ref{eq:9}).  (In computing $\chi^2$ we used
$\ell_{max}=1000$ for OMAP, $\ell_{max}=1500$ for CMAP and
$\ell_{max}=2800$ for Planck. However, the value of $\chi^2$ is
insensitive to $\ell_{max}$ since the errors in $C_\ell$ increase
exponentially at high multipoles.) The spatially flat models that we
have considered are defined by $7$ parameters, thus a $95\%$ rejection
of a model requires a $\chi^2$ of $\approx 14$. This is similar, by
construction, to the values of $\chi^2$ listed for OMAP. For CMAP, and
particularly, Planck, the values of $\chi^2$ are much larger,
demonstrating their superior ability to distinguish between
cosmological models. An experiment such as Planck can distinguish
between two $C_\ell$ curves at a high significance level even if the
differences between them are much smaller than the cosmic variance; in
fact the power spectra and residuals for a pair of models with $\chi^2
\approx 14$ would be indistinguishable plotted on the scales of Figure
\ref{figure11}.

Small, correlated errors in $C_\ell$ can thus introduce biases in
estimates of cosmological parameters. The degree of bias depends on
the size of the errors and the extent to which the correlated
component resembles the shape of the derivatives of $C_\ell$ with
respect to cosmological parameters.  The analysis of correlated errors
in $C_\ell$ can therefore be viewed as another aspect of the analysis
of degeneracies in cosmological parameter estimation.  There are many
possible sources of correlated errors in $C_\ell$. The most obvious
are errors caused by inaccurate subtraction of emission from the
Galaxy and extragalactic point sources. However, there are many other
more instrument specific sources of error, for example, in a
`total-power' experiment like Planck, the scanning strategy together
with instrumental `$1/f$' noise can introduce systematic errors in
maps of the CMB (sometime called `striping', Wright 1996, Bersanelli
\etal 1996, Tegmark 1997a, Delabrouille 1998). The resulting effect of
scanning on $C_\ell$ depends on the details of the experiment and
map-making methods, but for a fairly wide class of possibilities leads
to a contribution to $C_\ell$ that varies as $1/\ell$ (Tegmark 1997a).
The scanning errors in the CMB maps should be much smaller in a
differential experiment, such as MAP.  Fortunately, foreground and
scanning errors are unlikely to produce wiggles in $C_\ell$ that can
distort or mimic the Doppler peak structure at high
multipoles. Instead, we would expect smoothly varying errors to
correlate with the low-order principal components.

\begin{figure}

\vskip 3.0 truein

\includegraphics{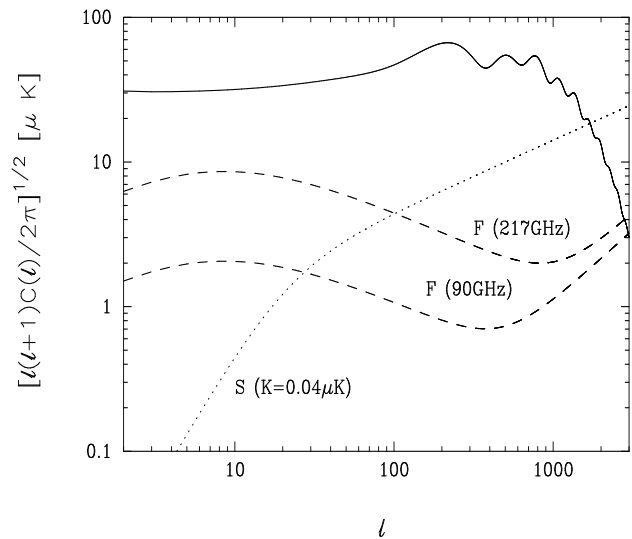}

\caption{The solid line shows the CMB power spectrum of the
spatially flat target model plotted with a logarithmic ordinate. The
two dashed lines show the power spectra of the foreground model
of equation (\ref{eq:32}) with the coefficients $A$ and $B$
set to the values appropriate for foregrounds at $90$ and $217$ GHz
as estimated by Bouchet \etal (1997). The dotted line shows the
scanning error  model of equation (\ref{eq:33}) with $\ell_K = 20$ and
$K = 0.04 \mu K$.}
\label{figure12}
\end{figure}

To illustrate the effects of correlated errors we have constructed the
following simple models
\begin{eqnarray}
 {\ell(\ell+1){\Delta C^F_\ell}\over 2 \pi}
 &=&  A^2 \ell (\ell+1) {1000 \over ( 5 + \ell)^3}  + B^2 {\ell (\ell+1) \over 1000^2},\label{eq:32}
\end{eqnarray}
and 
\begin{eqnarray}
 {\ell(\ell+1){\Delta C^S_\ell} \over 2 \pi}
&= &  K^2  {\ell_K \ell^3(\ell + 1) \over (\ell^2 + \ell_K^2)^{3/2}}\, .  \label{eq:33} 
\end{eqnarray}
The first model (which we denote `F') is designed to mimic the power
spectrum of contaminating foregrounds. The first term is a rough
approximation to the power spectrum of the major sources of Galactic
emission, free-free, synchrotron and dust emission (Low and Cutri
1994; Guarini, Melchiorri and Melchiorri 1995, Tegmark and Efstathiou
1996, Bouchet \etal 1997) . The second term represents the white noise from 
unresolved extragalactic point sources.  The analysis of Bouchet
\etal (1997) suggests the following values for $A$ and $B$ (in $\mu
K$) at $90$GHz and $217$GHz: 
\beglet
\begin{eqnarray}
&& A  = 0.36 , \qquad  B =  1.07, \qquad (90 {\rm GHz}),   \label{eq:34a}\\
&& A  =   1.5 , \qquad  B =  1.39, \qquad (217 {\rm GHz}).  \label{eq:34b}
\end{eqnarray}   
\endlet
The values of $A$ in (\ref{eq:34a},\ref{eq:34b}) are for the cleanest
regions of the sky. The second model (which we denote `S') is based on
the analysis of Tegmark (1997a) and is designed to approximate
correlated scanning errors. Values for the constants $K$ and $\ell_K$
are extremely sensitive to the experimental details but plausible
values for a `total-power' experiment may be $\ell_K \approx 20$ with
$K$ in the range $0.004$ --- $0.04\mu K$ (see Tegmark 1997a,
figure~\ref{figure10}). The models of equations (\ref{eq:32}) and
(\ref{eq:33}) are illustrated in Figure \ref{figure12}.

\begin{figure}
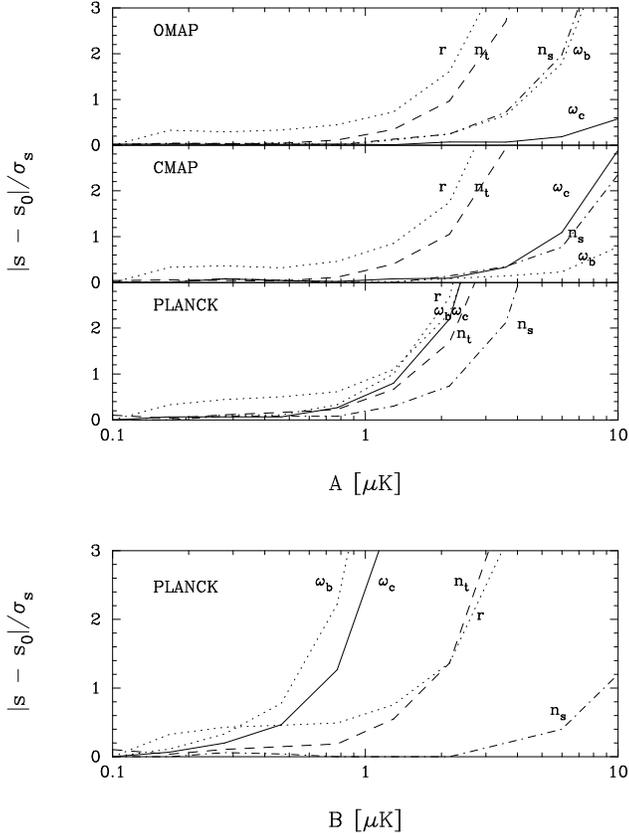


\vskip 4.5 truein

\includegraphics{pgfig13a.ps}

\includegraphics{pgfig13b.ps}

\caption{The absolute bias of various cosmological parameters
divided by their standard errors plotted against the parameter
$A$ representing Galactic foregrounds, and $B$ representing point
sources (equation \ref{eq:32}).}
\label{figure13}
\end{figure}

\begin{figure}

\vskip 3.0 truein

\includegraphics{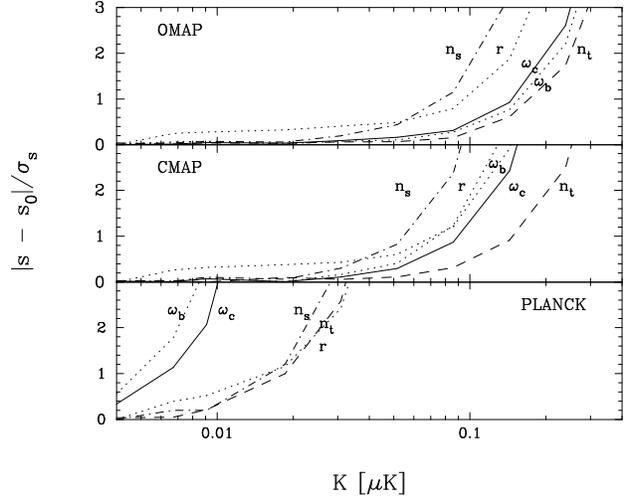}

\caption{The absolute bias of various cosmological parameters
divided by their standard errors plotted against the parameter
$K$ representing scanning errors according to equation \ref{eq:33}.}
\label{figure14}
\end{figure}

Figures \ref{figure13} and \ref{figure14} shows the biases in a  number of cosmological parameters
for various choices of the parameters $A$, $B$ and $K$. To compute
these figures, we have generated Monte-Carlo realizations of the
CMB power spectrum, to which we have added the systematic error
models of equations (\ref{eq:32}) and (\ref{eq:33}). We then compute values of the
cosmological parameters by fitting the first order expansion to
$C_\ell$ (equation \ref{eq:13}) as in the computations of Figures \ref{figure7}a and \ref{figure7}b.
The bias in each parameter has been divided by the 
expected standard deviation in each parameter to illustrate when
systematic errors  become larger than the random errors.  The results
are encouraging. For OMAP
and CMAP, the tensor parameters are significantly affected for values of $A$ 
$\simgt 1 \muK$ and the remaining parameters for values $A \simgt 3
\muK$. OMAP and CMAP are unaffected by  point source
contributions with $B \simgt 10 \muK$ because they
are insensitive to the power spectrum at high  multipoles where
point sources dominate. For Planck, the systematic biases become
significant for values of $A \simgt 1\muK$ and 
for $B \simgt 0.5 \muK$. Our analysis therefore
suggests that at the highest frequencies, and in clean regions of the
sky,  MAP should provide unbiased estimates of cosmological parameters
without any correction for foregrounds. For Planck,  removal of 
foregrounds is likely to be required to achieve unbiased results.
However, with the wide frequency coverage and high
sensitivity of Planck it should be possible to remove foreground
contributions to $C_\ell$ to an accuracy that is well below the
cosmic variance errors (Tegmark and Efstathiou 1996). Even without
any frequency information, it would be possible to test for biases
by including parametric models of the systematic errors
in the likelihood analysis.

Figure \ref{figure13} shows that for Planck, low-level scanning errors
could potentially lead to biases in cosmological parameters. These
biases could probably be removed adequately by including a parametric
model of the scanning errors derived from simulations such as those
described by Delabrouille (1998). The example of Figure \ref{figure13}
does, however, show that scanning errors may be a serious concern for
the analysis of Planck data.

\section{External Constraints and Parameter Degeneracies}\label{sec:6}

In practice, observations of the CMB anisotropies will be supplemented
by other astronomical observations which can be used to remove some of
the parameter degeneracies.  Some of these are described in this
Section. Indeed, such constraints have already been applied in some
analysis of cosmological parameters. For example, Webster \etal (1998)
describe joint likelihood analysis of CMB measurements and
observations of large-scale structure. White (1998) and Tegmark \etal
(1998) \ describe how CMB measurements can be combined with distances
derived from Type 1a supernovae ({\it e.g.} Perlmutter \etal 1997,
1998) to remove the geometrical degeneracy.

\subsection{Constraints on the Hubble Constant}\label{sec:6.1}

If the matter content parameters $\omega_m$ and $\omega_D$ are well
constrained by CMB observations, a constraint on the Hubble constant
converts directly into a constraint on $\Omega_\Lambda$.  For our
spatially flat target model, the constraint is given by equation (\ref{eq:12})
and generally,
\begin{equation}
\delta \Omega_\Lambda \approx 2 {\delta H_0 \over H_0}.\label{eq:35}
\end{equation}
A determination of $\Omega_\Lambda$ to an accuracy of 
$\sim 0.1$ should be possible if the 
systematic and random errors on the Hubble constant can be reduced
to  $\simlt 5 \%$, as seems feasible (see {\it
e.g.} Freedman \etal 1998).

\subsection{Constraints on the age of the Universe}\label{sec:6.2}

The age of the Universe is give by the integral,
\begin{equation}
 t_0 = 9.8\;{\rm Gy} \int_0^1 {a \;da \over [ \omega_m a + \omega_ka^2 
+ \omega_\Lambda a^4]^{1/2}} \label{eq:36}
\end{equation}
(see {\it e.g.} Peebles, 1993, equ 13.9). If the matter content
$\omega_m$ is well constrained by the CMB anisotropies, then limits
on the age of the Universe can be used to constrain the geometrical
degeneracy in the $\omega_\Lambda$ -- $\omega_k$ plane. This is 
illustrated in Figure \ref{figure14}. However, since the slope
of the  geometrical lines are so steep, the limits on
$\omega_\Lambda$ are sensitive to the parameters of the
target model. In the example
shown in Figure \ref{figure14}, an age constraint of $t_0 \simgt 14$ Gyr ({\it e.g.}
Chaboyer \etal 1996) would lead to a relatively weak limit
of $\omega_\Lambda \simlt 0.7$ in a spatially flat universe
(${\cal R}=2$), but to tighter limits in a universe with a 
larger value of $\omega_k$.

\begin{figure}

\vskip 3.0 truein

\includegraphics{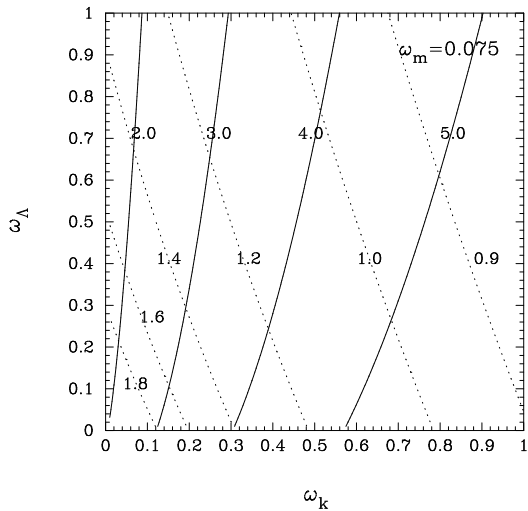}

\caption{Analogue of Figure \ref{figure1}b for a model with $\omega_m = 0.075$.
The  solid line show constant values of ${\cal R}$ in the $\omega_\Lambda$--
$\omega_k$ plane and the dashed lines show lines of contours of
equal age listed in 10 Gyr units.}
\label{figure15}
\end{figure}

\subsection{Constraints on Large-Scale Clustering}\label{sec:6.3}

Measurements of galaxy clustering can provide constraints on the shape
and amplitude of the galaxy power spectrum. These are often
characterised by the {\it rms} amplitude of the galaxy fluctuations in
spheres of radius $8 \hmpc$, $(\sigma_8)_g$, and a shape parameter
$\Gamma$ (see Efstathiou, Bond and White 1992). When combined with CMB
anisotropy measurements, observation of galaxy clustering can
constrain the distribution of mass fluctuations relative to galaxies.
This information, together with constraints on redshift space
anisotropies and cosmic velocity fields, can be used to improve the
estimates of many of the cosmological parameters described here and
including the epoch of reionization (see for example Tegmark
1997b). In the future, observations of weak gravitational lensing may
provide more direct estimates of the matter power spectrum (Blandford
\etal 1991, Miralda-Escude 1991, Kaiser 1992, see Seljak 1997b for a
recent discussion).  If we restrict to CDM models with low baryon
fractions, then the power spectrum shape parameter $\Gamma$ is related
to the matter density according to $\Gamma \approx \Omega_m h =
\omega_m/h$ (see Eisenstein and Hu, 1998, equations 30 and 31 for a
fitting function for $\Gamma$ that includes the baryon density and
associated scale dependence). A measurement of $\Gamma$ from
large-scale structure measurements can thus be combined with an
accurate determination of $\omega_m$ from the CMB anisotropies to
yield an estimate of the Hubble constant, and hence to constrain the
cosmological constant as described in Section \ref{sec:6.1}.

\subsection{Constraints on the Geometry from Type 1a Supernovae}\label{sec:6.4}

The distances of Type 1a supernovae at redshifts $z \simgt 0.5$ are
already providing strong constraints on the geometry of the Universe
({\it e.g.} Perlmutter \etal 1997, 1998).  Such measurements provide a
potentially powerful method of removing the geometrical degeneracy in
the CMB measurements. This is illustrated by Figure \ref{figure15},
which shows the CMB geometrical degeneracy lines in the
$\Omega_\Lambda$---$\Omega_k$ plane plotted as in Figure
\ref{figure1}a.  The dashed lines show constant values of the
luminosity distance at $z=0.5$\footnote{This is approximately the
median redshift of the distant supernovae samples, (Ellis, private
communication).}, approximating the slope of the error ellipses in
this plane derived from supernovae measurements. To first order,
standard candles such as Type 1a supernovae yield near degenerate
constraints along the dashed lines. As has been noted as well by White
(1998), Tegmark \etal (1998) and others, these intersect almost
orthogonally with the CMB geometrical degeneracy lines.  Type 1a
supernovae thus offer an extremely powerful way of breaking the
geometrical degeneracy, provided systematic errors (which move the
supernova maximum likelihood almost along the geometrical degeneracy
lines) can be shown to be small.

\begin{figure}

\vskip 3.0 truein

\includegraphics{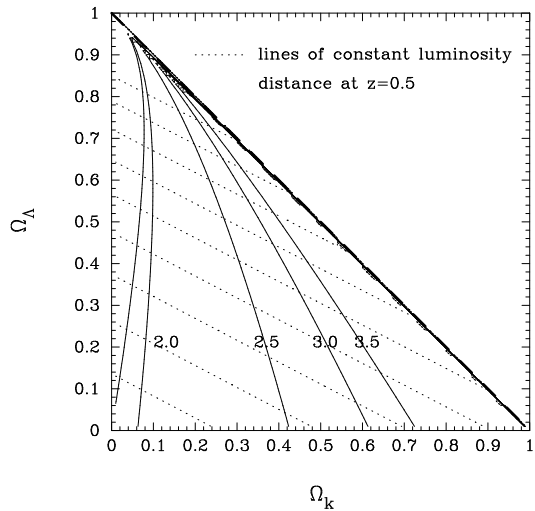}

\caption{The solid lines show the degeneracy lines of constant ${\cal
R}$ as plotted in Figure \ref{figure1}a. The dashed lines show lines
of constant luminosity distance at a redshift of $z=0.5$ in steps of
$50 \hmpc$, extending from $1700\hmpc$ at the bottom of the figure to
$2100\hmpc$ at the top.}
\label{figure16}
\end{figure}

\section{Conclusions}\label{sec:7}

High precision observations of CMB anisotropies promise a revolution
in our knowledge of fundamental cosmological parameters. In this paper
we have analysed the degeneracies between cosmological parameters
determined from CMB experiments, using the MAP and Planck satellites
as an indication of what might be achieved in the next decade. Our
conclusions are as follows:

\noindent
[1] The geometrical degeneracy between $\Omega_\Lambda$ and $\Omega_k$
for models with identical matter content, $\omega_b$, $\omega_c$,
$\dots$ is nearly exact and cannot be broken from observations of the
linear CMB anisotropies alone. Breaking of this degeneracy requires
additional data, such as distances to Type 1a supenovae, or the
detection of small non-linear effects such as gravitational lensing of
the CMB.

\noindent
[2] Small numerical errors in calculating the CMB power spectrum $C_\ell$
can accumulate in computations of the Fisher matrix leading to
spurious breaking of near exact degeneracies. These numerical errors
can be avoided by rotating to new variables defined by the
degeneracies as described in Section \ref{sec:3.3}.

\noindent
[3] The Fisher matrix for a particular experiment defines a set of
principal components, {\it i.e.} orthogonal linear combinations of the
cosmological parameters $s_i$ defining the theoretical CMB power
spectrum $C_\ell$. Restricting to a spatially flat universe and an
idealised set of $7$ cosmological parameters, we find that most of the
variance in estimates of the $s_i$ from the MAP satellite comes from
the two lowest order ({\it i.e.} least well determined) principal
components.

\noindent
[4] The degeneracies between poorly determined and well determined
parameters limits the accuracy of the Fisher matrix.  An exact
likelihood analysis for a MAP-type experiment shows that the Fisher
matrix can overestimate the errors on even apparently well determined
parameters such as $\omega_b$ by factors of two or more.

\noindent
[5] For a Planck-type experiment, the principal components are almost
identical to physical variables. The five highest order principal
components couple strongly to $\omega_b$, $\omega_c$, $Q$, $n_s$ and
$\omega_\Lambda$ respectively. The two lowest order components couple
strongly to the parameters $r$ and $n_t$ defining the tensor 
component.

\noindent
[6] For a MAP-type experiment, the errors on the matter densities
$\omega_b$ and $\omega_c$ are sensitive to theoretical constraints on
the tensor and scalar fluctuation spectra ($r$, $n_t$ and $n_s$) and
hence on the nature of the inflationary model. The variances on
$\omega_b$ and $\omega_c$ from a Planck-type experiment are
insensitive to the details of inflation, provided the tensor and
scalar fluctuation spectra are characterised by weakly scale-dependent
power-laws.

\noindent
[7] The lowest order principal components for an experiment sampling low
multipoles ($\ell \simlt 300$ in a spatially flat universe) can be
understood in terms of the position and height of the first Doppler
peak.

\noindent
[8] Correlated errors in $C_\ell$ that are much smaller than the
errors from cosmic variance can introduce significant biases in the
estimates cosmological parameters. We have analysed simple models for
correlated errors in $C_\ell$ arising from Galactic foregrounds and
extragalactic point source and from scanning errors.  If our simple
models are correct, foregrounds should not introduce large biases in
cosmological parameters estimated from CMB measurements at frequencies
in the range $\approx 100$--$200$GHz.  However, there may be
foreground components that are not included in our model and that
dominate in this frequency range ({\it e.g.} Leitch \etal 1997). In
this case, subtraction of the frequency dependent foregrounds from the
primary anisotropies to high accuracy may be necessary to derive
cosmological parameter estimates from the CMB. Scanning errors are
another potential source of biases in cosmological parameter
estimates. Our models suggest that scanning errors may be problematic
for Planck and a more detailed analysis is evidently required before
the scanning strategy is frozen.

\noindent
[9] Observations of CMB anisotropies can be supplemented with more
conventional astronomical measurements, such as the large-scale mass
distribution, estimates of the age of the Universe and distances to
Type 1a supernovae, to break the geometrical degeneracy.

\section*{Acknowledgements}
We thank Max Tegmark and Francois Bouchet for useful discussions. GPE thanks
PPARC for the award of a Senior Fellowship. JRB was supported by a 
Canadian Institute for Advanced Research Fellowship.

\end{document}